\documentstyle[pra,aps]{revtex}
\input epsfig.sty

\newcommand{\ra}{\rangle} 
\newcommand{\ba}{\begin{eqnarray}} 
\newcommand{\ea}{\end{eqnarray}} 
\newcommand{\be}{\begin{equation}} 
\newcommand{\ee}{\end{equation}} 
\newcommand{\bea}{\begin{eqnarray}} 
\newcommand{\eea}{\end{eqnarray}}

\newtheorem{lemm}{Lemma}

\newtheorem{theo}[lemm]{Theorem} 
 
\def\CC{{\rm\kern.24em \vrule width.04em height1.46ex depth-.07ex \kern-.30em C}} 
\begin{document}

\title{Encoded Universality from a Single Physical Interaction} 
 
\author{
J. Kempe,$^{1,2}$, 
D.  Bacon,$^{1,3}$ D. P. DiVincenzo,$^4$ and 
K. B. Whaley$^{1}$ 
} 
 
\address{
Departments of Chemistry$^1$, Mathematics$^2$ and 
Physics$^3$, University of California, Berkeley 94270 \\ 
$^4$IBM Research Division, T.J. Watson Research Center, Yorktown Heights, New York 10598}

\maketitle 

\begin{abstract} 

We present a theoretical analysis of the paradigm of encoded universality,
using a Lie algebraic analysis to derive specific conditions under which
physical interactions can provide universality.  We discuss the
significance of the tensor product structure in the quantum circuit model
and use this to define the conjoining of encoded qudits. The construction
of encoded gates between conjoined qudits is discussed in detail.  We
illustrate the general procedures with several examples from exchange-only
quantum computation.  In particular, we extend our earlier results showing
universality with the isotropic exchange interaction to the derivation of
encoded universality with the anisotropic exchange interaction, i.e., to
the XY model.  In this case the minimal encoding for universality is into
qutrits rather than into qubits as was the case for isotropic (Heisenberg)
exchange.  We also address issues of fault-tolerance, leakage and
correction of encoded qudits.

\end{abstract} 


\section{Introduction} \label{intro}

Many machines have been invented to perform certain specific computational 
tasks.  However in the language of computer science, these machines are not 
necessarily {\em computers}, meaning  
that they can not perform {\em all possible} computational tasks, i.e., they are not 
{\em universal}.  What is considered as {\em all possible computations} depends on 
the underlying laws that govern the machines we use.  In the classical world we 
can cast any computational task into a {\em boolean function} of the input 
(in bits).  The evaluation of a {\em computable} function can then be reduced to a 
sequence of elementary logical operations acting on a few bits each: {\em Classical 
circuits} can be build from a {\em universal gate set}.  The computational task of a quantum computer is to accurately perform {\em any unitary} operation on the input (qubits). 
These unitaries will have to be constructed from the basic units the computer 
will be equipped with, i.e., from a {\em  universal quantum gate set}. Moreover these realistic 
quantum gates have to be physically realizable - they should involve only a few 
qubits at a time, for example. 
 
Research efforts in recent years has shown that realistic {\em universal quantum gate sets} exist 
and that uniform {\em quantum circuits} can be built. A physical realization 
 needs to implement such a universal set in order 
to be considered a universal quantum computer.  Some of 
the gates in such a set are easier to implement than others. In 
several proposals some two-qubit gates are much easier to implement 
than one-qubit gates, yet the latter ones are needed to complete a universal 
set of gates. 
We have previously presented results on how to make do with a restricted 
set of gates that is not universal. This restricted set does not 
accurately enact every unitary operation on the system qubits, 
yet can nevertheless provide universality on a restricted subspace\cite{Bacon:99b,Kempe:00a}.  This approach, that we
have termed
``encoded universality'', employs the philosophy of {\em encoding} the information, in 
order  to make an a priori non-universal set of gates 
universal on the code\cite{Bacon:01}. This potentially reduces the experimental efforts of 
implementing gates that are hard to build. 
 
In this paper we show that our recent demonstration of universality 
from a single physical interaction achieved with the Heisenberg 
exchange  interaction\cite{Kempe:00a} can be extended to an 
anisotropic interaction, namely to the XY model.  This well-known 
Hamiltonian model is relevant to a number of two-dimensional condensed 
matter phenomena.  The result shown here indicates the generality of 
the approach of seeking universal encodings tailored to specific 
physical interactions \cite{Bacon:01}. 
 
A second aim of this paper is to expand the Lie algebraic analysis presented briefly in 
Ref.~\cite{Bacon:01}, to show specifically the conditions under which selected 
elements of the algebra generated by the physical interactions  can provide 
universality.  We distinguish between the general results of universality on a $d$-dimensional subspace provided by this approach\cite{Kempe:00a}, and a more 
practical attempt to develop encoded one- and two-qubit  gates that implicitly
imposes the standard model of arbitrary one-qubit rotations and CNOT gates on the encoded 
qubits\cite{Bacon:99b,Kempe:00a}.  The construction of these 
encoded gates is discussed in detail, including efficient implementation in
terms of discrete operations\cite{DiVincenzo:00a}.  We will also address the 
important issue 
of fault-tolerance and leakage detection and correction.  This will furnish a 
complete picture of how {\em fault-tolerance} meshes with encoded universality. 
 
The structure of this paper is as follows. We shall first provide a brief 
summary of the concept of universality within the quantum circuit model of 
computation in Section~\ref{hist}.  We then present the essentials of the Lie 
algebra  analysis underlying the problem of simulation of a quantum  circuit, 
and discuss the various levels of tensor product  structure that are required 
to implement a quantum circuit structure. The representation theoretic 
framework is presented in Section~\ref{formalism}, illustrated by a summary of 
the earlier result  for isotropic exchange 
\cite{Kempe:00a,DiVincenzo:00a,Bacon:01}. We also provide the mathematical 
connection to results for universal computation on decoherence-free 
subsystems \cite{Kempe:00a}.   The core of the paper follows in 
Section~\ref{anisotropic}, with a detailed  presentation of the derivation of 
universal encodings for the  anisotropic exchange interaction. Finally we 
address the issues of fault-tolerance and leakage (Sec.~\ref{leakage}) and 
close with discussion of some open topics (Sec.~\ref{outlook}). 
 
\section{Universality}\label{hist} 
 
 
\subsection{Quantum Circuits and Complexity}\label{qcm} 
 
To build a computer from a physical system for algorithmic 
applications means to provide its user with a set of ``units'' from 
which he or she can implement an algorithm. A classical computer has to evaluate boolean functions. A convenient model for classical 
computation is the {\em uniform} circuit-model.  The input to the 
circuit is given by a string of bits; wires carry them to elementary 
gates (such as $AND$ and $NOT$) which are again interconnected by wires. The computation carries the bits along the wires and executes the gates, when they are reached. The output wires carry the 
results. These circuits can be parametrized by the size of the 
problem, i.e. the number $n$ of input bits. Informally speaking such a 
family of circuits is {\em uniform} if given $n$, the corresponding circuit 
can be constructed efficiently (i.e. in polynomial time in 
$n$). This assures, that the complexity of a problem is not hidden 
into the design of the circuit. Such uniform circuits can be simulated 
at polynomial cost by a {\em universal circuit}, one of which is the 
circuit with the $NAND$ and the $COPY$ gates only 
\cite{Dewdney:89}. The notion of complexity defines which problems are hard and which are easy 
according to the amount of space and time (i.e. gates) a circuit uses to solve 
them. Although classical universal circuits may differ in the basic gate-sets they use, any universal circuit can simulate any other {\em efficiently}, i.e. with only  polynomial overhead in both space and time. Consequently the definition of complexity of a circuit can be made independently of the specifics of the gate-set used.

 
The Quantum Circuit Model (QCM) can be viewed as a transposition of 
the classical circuit model to the quantum world. A quantum computer implements unitary operations. The input is given 
by a string of {\em qubits}, connected to elementary gates. These 
gates constitute a universal set if the circuit can 
implement any unitary operation (and simulate any other circuit). A 
measurement on the output wires gives the result of the 
computation. Note that the QCM comes equipped with a canonical 
decomposition into a tensor product of small systems - the qubits (or 
qu-$d$its in the general case of $d$ dimensions).  This is of course not only necessary to 
define a basic unit (like a bit for a classical computer), but also to 
decompose a circuit into basic gates.  Both are indispensable for the 
uniformity of the circuit family: given a size of a problem (e.g. to 
factor a number given as a bitstring of length $n$) one should be able 
to efficiently (i.e. polynomial in $n$) find the necessary 
gate sequence. This would be impossible if the structure of the  
Hilbert space changed with increasing $n$. The tensor structure ensures that
this is not the case, and the QCM is thereby amenable to analysis in terms
of the {\em quantum complexity} of an algorithm\cite{Bernstein:97a}. 
 
We note here, as pointed out by many other 
authors\cite{Deutsch:85,Deutsch:89,Feynman:85a}, that all known physical 
models which can perform some form of quantum computation appear to be 
simulatable by the quantum circuit model.  We therefore take this 
empirically  verified hypothesis 
as a fundamental definition of  what it means to build a quantum 
computer.  Thus, for the issue of universality, some map between our 
physical system and the QCM must be established. In particular, 
one of the essential features of the QCM is a tensor product structure 
between different subsystems and therefore we shall impose this on our
physical system.

\subsection{Universal Gate Sets}\label{sec:gatesets} 
 
In both the classical and quantum regime, 
{\em universality results} have been established that ensure the 
equivalence of ostensibly different models of computation which may employ 
different sets of gates. 
Studies of universality in the QCM have culminated in the definition 
of a {\em universal gate set}.   A natural impulse is to make a close 
analogy to the classical case, i.e. to search for a finite set of 
gates that {\em exactly} implements every quantum circuit 
efficiently. However, these attempts are  doomed to failure: obviously a discrete set of gates can't be used to implement an arbitrary unitary operation {\em exactly}, since the set of unitary operations is continuous.
Powerful results have  nevertheless been established for the {\em approximate} 
simulation of quantum  circuits with a set of finite gates, and we 
will now briefly review the steps leading there. 
A more detailed account can be found 
in \cite{Nielsen:book}. 
 
 
To quantify the accuracy of an approximation, one needs a distance on 
matrices. On the finite $N$ dimensional matrix space, any metric is 
basically as good as any other (i.e. they differ by only 
constant factors). For example, we can use the trace-norm $d({\bf 
U},{\bf V})=\sqrt{1-\frac{1}{N}{\rm Re}\left[ {\rm Tr}({\bf 
U}^{\dagger }{\bf V})\right] }$, or the norm  $d({\bf U},{\bf 
V})=max_{|\Psi\rangle} \| ({\bf U}-{\bf V}) |\Psi\rangle \|$ 
($|\Psi\rangle$ is a normalized state). A matrix ${\bf V}$ is then 
said to approximate a transformation ${\bf U}$ to accuracy $\epsilon $ 
if $d({\bf U},{\bf V})\leq \epsilon $. 
 
The methodology employed in most universality proofs and
constructions starts with a set of {\em gates} and then operates within the 
{\em unitary group} generated by their repeated applications to show 
that the whole of $SU(N)$ is generated, where $N$ is the dimension of 
the space\footnote{A notable exception is the proof by DiVincenzo 
\cite{DiVincenzo:95a}}. 
Historically the first universal quantum gate sets were three-qubit 
gates (Deutsch \cite{Deutsch:89}). Earlier work by Toffoli 
\cite{Toffoli:80} on classical circuits has introduced a classical universal 
three-bit gate - the {\em Toffoli-gate} - which is reversible. Classical 
circuits based on the Toffoli-gate can simulate all other classical circuits 
efficiently and in a reversible way. Deutsch generalized this gate to 
a unitary three-qubit quantum gate and proved its universality.  In 
1994-96 several groups improved on Deutsch's circuits. 
DiVincenzo\cite{DiVincenzo:95a} demonstrated how to construct Deutsch's 
three qubit gate using only one and two qubit gates, and thus 
demonstrated that the physically unrealistic three qubit interactions 
of Deutsch are not necessary for universality.  Barenco {\em et 
al.}\cite{Barenco:95a} proved that  single qubit gates together with 
the controlled not (CNOT) allow one to {\em exactly} implement any 
unitary transformation.  This is now so often taken as the paradigm for
universal computation that we may refer to it as the ``standard model''.
Another memorable universality result is due independently to Deutsch, 
Barenco and Ekert \cite{Deutsch:95} and Lloyd \cite{Lloyd:95a} in 
1995. They prove using Lie-algebra arguments (see next section) that 
{\em almost any}  two-qubit unitary gate is universal. {\em Almost 
any} is to be taken in the generic sense: in the $16$-dimensional 
manifold of possible two-qubit unitary gates acting between $n$ 
qubits, the set of gates which {\it do not} yield a universal set of 
gates is a lower dimensional manifold. They implicitly assume that the 
generic gates (given by Hamiltonians $H_{ij}$) can be applied between 
all qubit pairs $i,j$ and in both directions (i.e. as $H_{ij}$ and 
$H_{ji}$), i.e. they assume that some sort of permutation (or 
exchange) gate is given. However many of the accessible interactions 
in physical systems are not of this sort - nature is often {\em 
non-generic}! 
 
To quantum compute in the presence of (inevitable) errors, quantum information 
has to be protected against noise. To this end several approaches have been 
developed, among which are the theory of quantum error correcting codes (QECC) 
\cite{Shor:96} and the theory of decoherence-free subspaces and systems (DFS) 
\cite{Zanardi:97c,Lidar:98a}.  In these approaches quantum information is 
encoded into parts of the system Hilbert space.  The theory of {\em 
fault-tolerant} quantum computation shows how to achieve universality in 
the presence of errors and imperfect quantum gates.  The basic idea 
is to work directly with encoded universal gates that act  
on an encoded space, whether this derives from a quantum error correcting code
or from a decoherence-free subspace or subsystem.
Considerations of fault-tolerance have led to a number of specific universal gate sets.  These gate sets are useful 
because it is known how to implement them 
within the context of fault-tolerant quantum error correction. 
We will mention only one of 
them here\cite{Boykin:99}, namely the set consisting of $\{H,P,CNOT,\pi/8\}$ where 
\be 
H=\frac{1}{\sqrt{2}}\left(\begin{array}{cc} 1 & 1\\ 1 & -1 \end{array} 
\right) \quad P=\left(\begin{array}{cc} 1 & 0\\ 0 & i \end{array} 
\right) \quad  \frac{\pi}{8}=\left(\begin{array}{cc} 1 & 0\\ 0 & 
e^{i\pi/4}\end{array} \right) \quad CNOT=\left(\begin{array}{cccc} 1 & 
0 & 0 & 0\\0 & 1 & 0 & 0\\0 & 0 & 0 & 1\\ 0 & 0 & 1 & 0 \end{array} 
\right). 
\ee 

For practical implementations, the degree of locality of the gates is
important.  For QECCs it has been shown how to implement such an 
encoded discrete universal set of gates using local gates on the physical 
qubits \cite{Shor:96}. These gates take encoded states to encoded states, but 
may take the state outside of the encoded subspace during the computation. 
In order to apply this approach, the speed of gate operations is critical: it has to be less than the error rate. 
In the theory of computation on DFSs, Bacon et al. \cite{Bacon:99b} (for a 
specific class of DFSs) and Kempe et al. \cite{Kempe:00a} (on a broader class 
of DFSs) have shown how compute on DFS with local Hamiltonians only. This work 
 shows that the two-body exchange Hamiltonian alone is sufficient to implement  universal computation on a decoherence-free subsystem. This demonstrates in 
particular, that there is an {\em encoding} of the quantum information into a 
subspace such that the exchange Hamiltonian - which is a priori not universal in the sense 
that  it can perform any unitary evolution on all qubits - does now {\em become} universal. 
We have termed this ``encoded universality''. Ideas about encoded universality in the 
context of physical realizations of a quantum computer have been presented in 
\cite{Bacon:01,DiVincenzo:00a,Kempe:ph,Bacon:PHD} and also implicitly in 
\cite{Viola:00a} in the context of dynamical noise suppression 
schemes (so called ``bang-bang'' control). 

\subsection{Lie Analysis of Intrinsic Physical Interactions}\label{Lie} 
 
Every quantum gate is generated by a Hamiltonian. It is very convenient to 
recast the universality results into a Hamiltonian language. This allows us to 
use Lie-algebra analysis to approach universality constructions.  Physically, 
the implementation of a unitary quantum gate will come about due to the implementation of some controllable Hamiltonian for a specific time. 
 
Instead of supposing that we are given a set of unitary gates, we thus rather assume that we have control over a set of Hamiltonians which we can apply to 
our system for a controllable amount of time.   Of 
course, there will be inaccuracies in the application of a Hamiltonian for 
any given exact time.  These inaccuracies, however, are 
exactly what fault-tolerant quantum computation is designed for. 
Consequently it is legitimate to proceed from the assumption  
that we can exactly implement a given Hamiltonian for a given amount of 
time.  The aim is then to a) reformulate the analogue of universal sets of 
unitary gates Hamiltonians formulation and b) to provide efficient finite 
time approximations that result in implementable sets of universal gate 
operations.
 
We can now ask what new gates (and new Hamiltonians) we can obtain by 
operating these given ones.  A general theoretical procedure to combine known gates to construct new
gates is well known, deriving from the Baker-Hausdorff-Campbell operator
expansion.  Accordingly, we can combine two Hamiltonians to construct new 
Hamiltonians using the following properties: 
\begin{eqnarray} 
\label{Trotter1} 
&&e^{i(\alpha \bf{A}+\beta \bf{B})}=\lim_{p\rightarrow \infty 
}(e^{i\alpha \bf{A}/p}e^{i\beta \bf{B}/p})^{p}  \\ 
&&e^{i[\bf{A},\bf{B}]}=\lim_{p\rightarrow \infty }(e^{-i\bf{A} 
\sqrt{p}}e^{i\bf{B}/\sqrt{p}}e^{i\bf{A}/\sqrt{p}}e^{-i\bf{B} 
\sqrt{p}})^{p}. 
\end{eqnarray} 
 
In practice we will truncate the infinite series to approximate the relevant operations. A detailed analysis of this is provided in \cite{Lloyd:95a,Kempe:00a}. As a consequence of these composition laws scalar multiples, sums and 
Lie-commutators $i[{\bf A},{\bf B}]$ can be obtained out of the given Hamiltonians). Note that these operations correspond to closing the set of allowed 
Hamiltonians as a {\em Lie-algebra}.
To approximate to a given accuracy, say $\exp (i(\alpha \bf{A}+\beta 
\bf{B}))$, by a sequence of gates with generators $\bf{A}$ and $ 
\bf{B}$ one just truncates Eq.~(\ref{Trotter1}) for large enough 
$p$. This allows us to approximate every unitary element in the {\em 
Lie-group} corresponding to the Lie-algebra generated by the given Hamiltonians using 
controllable coupling constants. The inverse of this statement is also 
true, namely all gates that we can build of a set of given interactions are generated by elements in the Lie-algebra.

It is now easy to define a {\em universal set of generators} $\sf H$ 
in this setting: $\sf 
H$ is universal if the Lie-algebra generated by the elements of $\sf 
H$ contains $su(N)$, where $N$ is the dimension of the underlying 
space and $su(N)$ is the Lie-algebra of the special unitary group. This implies that the set of {\em unitary} gates obtainable via 
successive application of {\em gates} deriving from $\sf H$ is dense 
in the group $SU(N)$. 
 
This general approach provides a systematic way to
construct universal sets of gates, but so far it says nothing about whether
or not the resulting gates can be implemented efficiently.  We remind the
reader that we are using the term ``efficiently'' in the quantitative sense
defined above, namely in terms of whether implementation can be made with
only polynomial overhead in the number of discrete gates required.  The
Trotter expansion, Eq.~(\ref{Trotter1}),  is guaranteed to be accurate in the
limit as the number of terms $p$ becomes very large, but clearly a critical
practical issue is how the truncation scales with the size of the quantum
circuit, $n$.  
Thus in order to actually use this approach to universality for realistic 
physical implementation, it is crucial to have bounds 
on the length of the gate sequences approximating a certain gate in terms of 
the desired accuracy for a circuit of given size. 
This is all the more important if one universal set is 
to be replaced by any other with only polynomial overhead in the number of 
gates applied, for otherwise the complexity classes would not be robust 
under the exchange of one set for another. The Solovay-Kitaev 
theorem\footnote{For details and proofs see \cite{Nielsen:book}} 
establishes the equivalence of universal sets, and provides bounds on 
the length of gate sequences for a desired accuracy of 
approximation. In short, this theorem states: 
\begin{quote} {\it Theorem 
(Solovay-Kitaev): ---} Given a set of gates that is dense in 
$SU(2^{k})$ and closed under Hermitian conjugation, any gate ${\bf U}$ 
in $SU(2^{k})$ can be approximated to an accuracy $\epsilon $ with a 
sequence of ${\rm poly}\left[ \log (1/\epsilon )\right] $ gates from 
the set. \end{quote} 
It follows that any quantum circuit (performing any desired unitary 
operation) can be simulated to arbitrary accuracy 
efficiently by a circuit with gates from  a universal set. In the context of a universal set of Hamiltonians  $\sf H$ it implies that all the gates in the Lie-group can be obtained efficiently by an appropriate choice of the control parameter. 
This powerful result lays open the route to searching for 
radically different universal sets of operators. 
 
\section{Formalism} 
\label{formalism} 
 
Designs for quantum computers can come from a variety of underlying physical 
systems.  A major challenge for future quantum engineers is to find the 
physical systems that allow for a universal set of gates. In all the physical 
implementations proposed only particular Hamiltonians can be turned on and off. 
As mentioned in Sec.~\ref{sec:gatesets}, generic two-body 
interactions can suffice to implement any unitary transformation. Nature, 
however, is not generic. Most interactions in real physical system 
possess symmetries that place these interactions 
squarely on a lower dimensional ``non-generic'' manifold where they are not
universal.  Thus, while allowing in principle for universality 
based entirely on two-body interactions, this approach does not 
provide a general prescription for universality given a specific 
set of intrinsic interactions. 
 
We propose a different approach here, namely to start from the natural 
interactions 
given by the physics of the proposed qubit system, and then to 
investigate the underlying potential of these interactions for 
``encoded universality''. The rationale behind this is that the given 
interactions - although {\em per se} not universal - become {\em 
universal within a subspace}. Encoding into this subspace allows to 
reduce the number of different Hamiltonians needed for universality, at 
the expense of spatial resources. This approach allows one to  take the natural physical constraints of a specific implementation into account.  
For a given physical system, the natural  interactions 
of choice can then be determined by  various factors, including their 
speed, the ease with which they can be implemented, any engineering or
material constraints, and their 
robustness towards decoherence processes. 
 
Here we detail this new paradigm of ``encoded universality''.  We start 
by reviewing the relevant results from our previous work addressing quantum
computation on 
decoherence-free subsystems and by summarizing the mathematical framework. 
We 
illustrate the general theory with a summary of our previous results on the 
isotropic exchange interaction \cite{Kempe:00a,DiVincenzo:00a,Bacon:01}, 
before presenting the new results for the anisotropic exchange interaction in
Section~\ref{anisotropic}.
 
\subsection{Representation Theoretic Framework} 
 
The general formalism employed to find universal encodings is largely 
related to the Lie-algebra framework for {\em noiseless} or
{\em decoherence-free} (DF-) subsystems\cite{Knill:00a}. 
We briefly review this framework here.

In the context of DF- subsystems \cite{Knill:00a} we are dealing with operators in the interaction 
Hamiltonian ${\bf H}_I=\sum_\alpha {\bf S}_\alpha \otimes {\bf B}_\alpha$. The Hamiltonian ${\bf H}_I$ describes the system-bath interaction which ultimately causes decoherence and corruption of the information stored in the system. The 
$S_\alpha$ (acting on the system) generate an {\em interaction algebra} ${\cal S}$, which is the algebra generated by the set
${\cal S}_1=\{I,S_1,S_2,\ldots \}$ by {\em 
linear combination} and operator {\em multiplication}. Because the
interaction Hamiltonian is Hermitian, this algebra is $\dagger$-closed. 
The algebra $\cal S$ splits  into irreducible components as a
consequence of the following general theorem from the representation
theory of $\dagger$-closed algebras (see, e.g., \cite{Landsman:98a}):

\begin{theo} \label{th:daggertheorem} 
Let ${\cal S}$ be a $\dagger$-closed algebra of operators acting 
on a space ${\cal H}_S$. Then the space is isomorphic to a direct 
sum 
\be \label{eq:spacesplit} 
{\cal H}_S \sim \sum_{J\in {\cal J}}\CC^{n_{J}}\otimes \CC^{d_{J}} 
\ee 
in such a way that in this representation $\cal S$ and the 
commutant ${\cal S}^\prime$ of $\cal S$ are decomposable
respectively, as 
\be \label{eq:algebrasplit} 
{\cal{S}}\cong \bigoplus_{J \in {\cal J}}  I_{n_J} \otimes 
M(\CC^{d_J}) \quad \quad {\cal{S}^\prime}\cong \bigoplus_{J \in 
{\cal J}}  M(\CC^{n_J}) \otimes I_{d_J} 
\ee. 
\end{theo} 
Here $M(\CC^n)$ means the set of {\em all} linear operators from 
$\CC^n$ to itself, and $J \in {\cal J}$ runs over all the 
irreducible representations (of dimension $d_J$ and with degeneracy 
$n_J$) of $\cal S$. The commutant ${\cal S}^\prime$ of 
$\cal S$ is the space of all operators commuting with $\cal 
S$. It is itself an {\em associative algebra}, i.e. an algebra closed 
under linear combination and multiplication. 
States encoded in $\CC^{n_J}$ are completely immune to the interaction 
with the bath, because the interaction algebra $\cal S$ acts only 
on its ``co-factor'' $\CC^{d_J}$. Thus $\CC^{n_J}$ is a {\em 
noiseless} or {\em decoherence-free} subsystem where information is intrinsically stabilized 
against the effects of the interaction with the bath, with no need for 
corrective action. 
It is important to note that the {\em Lie-algebra} generated by the ${\bf 
S}_\alpha$ via {\em linear combination} and {\em Lie-commutator} is different 
from the interaction algebra ${\cal S}$. This point will become important 
later. 
 
One of the most important physical models of decoherence, which allows for decoherence-free subspaces and subsystems, is the model of so called {\em collective decoherence}.  In the case of collective decoherence on $n$ qubits the {\em 
interaction algebra} $\cal S$ is spanned by ${\bf 
S}_\alpha=\sum_{i=1}^n \sigma_\alpha^i$, with $\alpha=x,y,z$. This algebra is well known in the quantum mechanical literature, it represents a (reducible) representation of the Lie-algebra of $su(2)$. Its block-structure according to Theorem \ref{th:daggertheorem} allows to find the decoherence-free encoding of the information.

Yet this encoding alone leads  only half the way towards decoherence-free quantum computation. In order to harness DF-subsystems as quantum computers it has to be shown that one can perform universal computation {\em on DFSs} in a fault-tolerant way using local gates on the physical qubits. Inspecting the statement of Theorem \ref{th:daggertheorem} more closely it is not hard to see that operations that take DF-states to DF-states must be generated by Hamiltonians which lie in the commutant ${\cal S}^\prime$ of the interaction algebra $\cal S$. For universal computation within a DF-subsystem we need to identify local gates in this commutant (these will be the Hamiltonians we can control) and to show that the Lie-algebra generated by these Hamiltonians contains $su(n_J)$ on each cofactor $\CC^{n_J}$.

In our work on fault-tolerant universal computation on 
decoherence-free subsystems we showed \cite{Bacon:99b,Kempe:00a} that 
the exchange interaction alone generates the Lie-algebra $su$ on 
every such DF-subsystem. This result was obtained for any set of $n$ qubits, 
$n \geq 3$ and settled the question of universal fault-tolerant computation 
on decoherence free subsystems with local interactions.
 
Let us now switch viewpoints from consideration of the subsystem to 
consideration of the interaction. Our result \cite{Bacon:99b,Kempe:00a} then 
implies that if we restrict the Hilbert space of the system to the space 
of the subsystem only, the exchange interaction {\em alone} is 
universal. In other words, the encoding into subsystems provided a way 
to make the exchange universal, meaning that the set of exchange 
interactions ${\sf E}=\{{\bf E_{ij}}:\,1 \leq i < j \leq n\}$ on $n$ qubits 
is a {\em universal set of generators over a subspace} according to the 
definition in Sec. \ref{Lie}! 
We point out that while this result was achieved specifically for the 
decoherence-free subsystems that are defined by resistance to collective 
decoherence, the decoherence-free properties of the subsystems are not 
required for the universality result.  

The scenario of {\em encoded universality} is in a sense the ``inverse'' problem to universal computation on DFSs. In contrast to the DFS situation, we are 
now presented with a set of given interactions ${\sf H}$ on the system Hilbert 
space ${\cal H}_S$. The operations we can implement from this set 
generate a {\em Lie-algebra ${\cal L}({\sf H})$}. The {\em 
Lie-group} generated by this Lie algebra (via ``exponentiation'') is a 
subgroup of the  unitary group $U$ and thus is a 
compact Lie group. Compact Lie groups, along with the Lie algebra 
which generates the Lie group, are completely reducible\footnote{A 
representation of a Lie algebra is irreducible if the action of the 
representation on its vector space does not possess an invariant subspace.} 
and hence the Lie-algebra splits into a direct sum of irreducible 
representations (irreps) 
\begin{equation}\label{eq:Liesplit2} 
{\cal{L}}({\sf H}) \cong \bigoplus_{J \in {\cal J}} 
{\cal{L}}_{J}(n_J) \otimes I_{d_J} 
\end{equation} 
over the state space 
\be \label{eq:spacesplit2} 
{\cal H}_S \sim \sum_{J\in {\cal J}}\CC^{n_{J}}\otimes \CC^{d_{J}}. 
\ee 
Here ${\cal{L}}_{J}$ is the $J$th irrep of ${\cal{L}}({\sf H})$.  
This decomposition is reminiscent of 
that of the commutant ${\cal S}^{\prime}$ in Eq. (\ref{eq:algebrasplit}),
and so we use the same variables for the dimension $n_J$ of the $J$th irrep 
of ${\cal {L}}_{J}$ and its degeneracy $d_{J}$.  (Note that these
variables are interchanged for the interaction algebra ${\cal S}$.)
 
The problem we face in order to make the set $\sf H$ universal can now be 
formalized as follows: 
\begin{center} 
{\quote \em Given the Lie algebra ${\cal L}({\sf H}) \cong 
\bigoplus_{J \in {\cal J}} {\cal{L}}_{J}(n_J) \otimes I_{d_J}$ 
find a component $\CC^{n_J}\otimes \CC^{d_J}$ of the state space over 
which ${\cal{L}}_{J}(n_J)$ contains $su(n_J)$.} 
\end{center} 
If this can be satisfied, then we have achieved universality on a subsystem.

The exchange-only universality theorem we proved in \cite{Kempe:00a} shows that the 
{\em Lie-algebra} generated by the exchange interaction
generates $su(n_J)$ on 
{\em each} component $\CC^{n_J}$. This implies that exchange lends 
itself for encoded universality. We can encode into any of these 
components (corresponding to subsystems in the DFS framework). The 
problem of the existence and structure of ``encoded universality'' 
with isotropic exchange is thus solved in the affirmative. 
 
Furthermore, we can immediately infer the coding efficiency of the ``encoded 
universality'' obtained from exchange. As shown, e.g., in 
\cite{Kempe:00a} the number of encoded qubits $k$ over the number of 
physical qubits $n$ approaches $1$ as $n \rightarrow \infty$. This means that for larger and larger $n$ the amount of redundancy in physical qubits that are needed to encode approaches zero. 
However, 
as was outlined in Sec. \ref{qcm}, we do have to introduce a tensor 
product structure via what we will call {\em conjoining} (see Sec. \ref{tensor}), and this will place a
constraint on the encoding efficiency. 

The tensor product structure is important in order to 
map the QCM onto our subspace
. 
The convenient way to proceed in construction of this structure
is to use ``clusters'' of subspaces over which exchange 
is universal.  These clusters may be delimited either by some natural length
scale in the 
physical system, or by an arbitrary cut-off. For instance the {\em 
smallest} possible block size consists of $3$ physical qubits (to 
encode two states - a logical qubit) (see Sec. \ref{exchange}).  
The exchange interaction allows us to perform operations both within and 
between the blocks.  The encoding efficiency with such a tensor product 
structure is then limited by the block size to a value less than unity, no 
matter how many qubits are employed.  Choosing a higher cut-off, {\em i.e.}, a larger block size does increase 
the coding efficiency, but this will always remain at a value less than
unity for finite block size.
We will give more details of the conjoining procedure in 
Sec. \ref{tensor}. 
 
We will briefly 
review some 
specific examples for isotropic exchange in Sec. \ref{exchange}. 
In Sec. \ref{anisotropic} we will show that a similar result 
also holds for the XY-interaction, {\em i.e.}, for an anisotropic exchange
interaction. 

\subsection{Interaction Algebra $\neq$ Lie-Algebra} 
 
The algebra representation Theorem 
\ref{th:daggertheorem} has given rise to some misunderstanding, which 
we now clarify. 
Let us look at the case of collective decoherence on $n$ qubits. The {\em 
interaction algebra} $\cal S$ is spanned by ${\bf 
S}_\alpha=\sum_{i=1}^n \sigma_\alpha^i$, with $\alpha=x,y,z$, via linear 
combination and {\em multiplication}. The {\em Lie-algebra} spanned by the 
${\bf S}_\alpha$  - we will denote it by ${\cal L}(S)$ - represents a 
(reducible) representation of the Lie-algebra $su(2)$.  This is 
{\em different} 
from $\cal S$, since it is obtained by linear combination and 
{\em taking commutators}, i.e. in general ${\cal L}(S) \subseteq \cal S$. 
However, for compact groups (and the $S_\alpha$ form a compact group), 
it is true that when expressed in the same basis as $\cal S$,
the {\em Lie-algebra} ${\cal L}(S)$ splits into
the same block-structure as that of Eq.~(\ref{eq:Liesplit2}). 
 
 
As we can easily infer from the DFS theorem  \cite{Zanardi:99d,Kempe:00a},
if the encoded operations are to preserve the information in the 
encoded subsystems, the gate Hamiltonians should be in the commutant 
${\cal S}^\prime$ of $\cal S$. It is exactly these operations 
in ${\cal S}^\prime$ that act entirely within the subsystems, as 
can be seen from Eq.~(\ref{eq:algebrasplit}). Yet it turns out that the commutant of ${\cal L}(S)$ and the commutant of $\cal S$ are the same; in both cases it consists of the set of elements that commute with all the ${\bf S}_\alpha$.
 
The good news is that instead of all $\cal S$, we only need to 
study the structure of the irreps of $su(2)$ as a {\em Lie-algebra} 
${\cal L}(S)$. It is very well known in quantum-mechanics 
since it is precisely the Lie algebra deriving from angular momentum. This will give us the noiseless or decoherence-free
subsystems, and also the commutant, ${\cal S}^\prime$.  
Note that the Lie algebra $su(2)$
arises specifically from the DFSs resulting from the collective decoherence model.
Other decoherence models will yield different DFSs that may be
defined by different Lie algebras.
 
As is well known from the representation theory of $su(2)$ (see e.g. \cite{Cornwell:84a}) the commutant of ${\cal L}(S)$ is 
related to the {\em symmetric group} (permutation group) $S_n$ 
\cite{Zanardi:99d}. The {\em natural representation} of $S_n$ on $n$ qubits is 
the set of operators that permutes the qubits, i.e. if $\pi \in S_n$, then $\pi$
acts on basis states as \be \label{eq:naturalS_n} \pi: \,|i_1\ra  |i_2\ra 
\ldots  |i_n\ra \longrightarrow |i_{\pi(1)}\ra  |i_{\pi(2)}\ra  \ldots 
|i_{\pi(n)}\ra \quad \quad i_k \in \{0,1\} \ee 
Clearly the elements of $S_n$ commute with the permutation-invariant 
${\bf S}_\alpha$. But they - together with the associative algebra they generate -  also constitute the set of {\em all} elements that commute with them. In other words the commutant of ${\cal 
L}(S)$ (this is equal to ${\cal S}^\prime$, the commutant of 
${\cal S}$) is given by 
the algebra spanned by linear combinations  of elements of $S_n$ in its 
natural representation, Eq.~(\ref{eq:naturalS_n}). 
The {\em 
associative algebra} spanned by the permutation group $S_n$ gives the set of 
{\em all} operators on a subsystem, as can be seen from 
Eq.~(\ref{eq:algebrasplit}).
In particular, we can find a set of generators for {\em universal 
computation} on subsystems among its elements. This means that for universal 
fault-tolerant computation all we have to do is to implement the 
permutation-group. 
It is also known that the permutation-group $S_n$ is generated by 
transpositions $\tau_{ij}$ (which just permute two qubits). These 
transpositions can obviously be implemented by local interactions - 
just switch on the exchange Hamiltonian. One might be tempted to 
conclude that the question of universal computation with local gates 
is thus settled! 
 
Unfortunately this conclusion is false. The transpositions generate 
the permutation group $S_n$ via {\em multiplication}. But the 
composition laws for given operations do not generate {\em products} 
of allowed Hamiltonians. The allowed ways to compose operations 
from a set of basic Hamiltonians give only {\em linear combinations} 
and {\em commutators}, i.e. they close the {\em Lie-algebra} of the basic set 
(Sec. \ref{Lie}). In other words, if our basic 
interactions are the exchange (transposition) of two qubits, we can 
implement the {\em Lie-algebra} generated by them, but not the 
associative algebra. 
 
For universal  computation with exchange we had therefore to show 
explicitly 
\cite{Kempe:00a} that the Lie-algebra generated by the transpositions 
$\tau_{ij}$ generates the whole Lie-algebra of $su(n_J)$ on a 
subsystem $\CC^{n_J}$. 
 
In order to clarify  this it is useful to discuss an example.
Suppose that we are given a 
system of $n$ qubits upon which we can enact one of the three collective 
Hamiltonians 
\begin{equation} 
{\bf C}_\alpha = \sum_{i=1}^n \sigma_{\alpha}^i. 
\end{equation} 
The Lie algebra ${\cal A}$ generated by these interactions is simply that of $su(2)$: 
\begin{equation} 
[{\bf C}_\alpha, {\bf C}_\beta] = i \epsilon_{\alpha \beta \gamma} {\bf 
C}_\gamma. 
\end{equation} 
This Lie algebra will thus be reducible to a set of $n_J$ dimensional irreps of 
$su(2)$: 
\begin{equation} 
{\cal A} \cong \bigoplus_{J=0 (1/2)}^{n/2}  {\cal 
A}_J(n_J) \otimes I_{d_J}, 
\end{equation} 
where ${\cal A}_J(n_J)$ is the $n_J=2J+1$ dimensional irrep of $su(2)$ which appears 
with degeneracy $d_J$ in the decomposition.  It is important to realize that 
if a set of Hamiltonians represents an $n_J$ dimensional irrep of $su(2)$, then  
the {\em operational} power of these interactions is not more than 
the operational power of any other irrep of $su(2)$. In particular, it is
no more 
powerful than the two dimensional irrep of $su(2)$.  
An $n_J$ dimensional representation of $su(2)$ acts on a $n_J$ dimensional space, 
but does not enact $su(n_J)$ on this $n_J$ dimensional space. 
Yet if we look at the {\em associative } algebra spanned by the 
${\bf C}_\alpha$ and its decomposition into a block structure according to 
Theorem \ref{th:daggertheorem}, we see that on each co-factor of each block 
this algebra generates every possible matrix, and in particular, all 
matrices in $su(n_J)$ (it is the same associative algebra that we studied in 
the case of collective decoherence). The point is that the power of the associative algebra is nevertheless not accessible via the fundamental physical composition laws.
 

\subsection{Conjoining and the Tensor Product Nature of Computation} 
\label{tensor} 
 
The above summary of representation theory for Lie algebras has been made 
solely in terms of some abstract $d$-dimensional Hilbert space. We now admit 
two tensor product structures: first, on our physical system, and second, on 
the encoded qubits from which encoded universality will be constructed. 
 
 The first tensor product structure is simply that implied by our physical 
system and is usually forced on us by the locality of physics.  For 
example, if we are using the spins of single electrons on a quantum 
dot, the tensor product is just the natural one of these spin-qubits. 
The nature of this tensor 
product structure is also manifest in the set of interactions which will be 
present in the real world, and is represented in the Hamiltonians in 
$\sf H$ at our disposal.  This tensor product is not relevant to 
our discussion except in the context that we expect the physical 
interactions  we are dealing with to be {\em local} with respect to it. 
 
The second tensor product is the important tensor product relevant to 
building  a map between the given interactions and the QCM. As 
mentioned before {\em within} an {\em subspace for universality} there 
is a priori no tensor structure present. In order to build uniform 
quantum circuits for quantum computation we need to introduce a tensor 
structure at some point ( Sec. \ref{qcm}).  In particular we are 
seeking to use a certain set of interactions to simulate a QCM and 
therefore we require that there be a tensor product structure 
corresponding to the the tensor product structure in this model. 
 
In order to define a tensor product structure, one must be able to identify 
subsystems with a tensor product structure between them.  The exact method for 
establishing where this tensor product structure comes into play is arbitrary. 
From a physical perspective one is motivated to search for {\em small 
encodings}, i.e., we seek to minimize the number of physical qubits used to 
encode a logical qubit (or qu{\em dit} if d-dimensional). The tensor product structure will then be induced between these 
encoded qubits (qudits) by separating them into blocks (e.g., of nearest neighbors), 
with the defining property that single encoded-qudit operations are possible 
within each such block. However the cut-off is in principle largely 
arbitrary and so our ultimate choice will depend on the balance between
a desired coding efficiency 
versus the complexity of gate sequences required for this.
These blocks of encoded qubit (or qu{\em dit} if 
d-dimensional) systems will form the basic factors of our ultimate 
tensor structure. We term this process {\em ``conjoining''} the 
encoded qudits\cite{Kempe:00a}. Within each block we can implement any encoded 
operation as a consequence of our universality proofs. 
 
A subtlety arises when we want to perform encoded operations {\em 
between} the blocks. This has to be addressed separately from the operations
within a single block. In the case 
of the exchange interaction we have the property that the space formed 
by the two separate encoded qudits is itself part of an encoded higher 
dimensional system. This is a consequence of the structure of the 
subsystems, as shown explicitly in the proof in \cite{Kempe:00a}. This means 
that the exchange interaction implements any operation within this bigger 
subsystem, including
in particular, any operation we want to perform between the 
blocks. A similar statement is true in the case of the XY-interaction. 
 
Finally, given a certain encoded universality construction, the issues 
of preparation and measurement on the encoded information must be 
established.  Generally speaking, some method of efficiently 
generating states with a large overlap of the encoded information must 
be possible. Similarly, a method for extracting some bit of 
information from the encoded information must also be available. 
These are important issues which need to be addressed within the 
physical feasibility of a given encoded universality construction. 

Finally we note that the formalism we have developed here works only for a 
quantum circuit where measurement is not used as a fundamental process 
for enacting unitary evolution.  Recently it has been demonstrated 
that universal quantum computing can be achieved using only 
measurements on a particular prepared entangled 
state\cite{Raussendorf:01a}.  Our formalism does not deal with 
measurements used to construct approximate unitary evolution. 
 
\subsection{Example - Universality from Isotropic Exchange Hamiltonian} 
\label{exchange} 
We will illustrate the theory of encoded universality with an example of 
our previous results on the isotropic exchange interaction\cite{Kempe:00a}. 
 The most general form of the Heisenberg exchange interaction between 
two spins is  the fully anisotropic spin-spin coupling 
\begin{equation} 
{\bf H}_{ij}=J^{X}_{ij} {  \sigma}_x^i {  \sigma}_x^j+J^Y_{ij} { 
\sigma}_y^i {  \sigma}_y^j +J^{Z}_{ij}{  \sigma}_z^i {  \sigma}_z^j. 
\label{eq:hei_anis} 
\end{equation} 
In the isotropic limit the exchange 
couplings between a given $ij$ pair are all equal, {\it i.e.}, 
$J^{X}_{ij} = J^{Y}_{ij} = J^{Z}_{ij} \equiv J_{ij}$, so that 
we are then dealing with the Heisenberg Hamiltonian 
\begin{equation} 
{\bf H}_{\rm Hei} =  \sum_{i \neq j} J_{ij} {\bf E}_{ij}. 
\label{eq:hei_iso} 
\end{equation} 
To show how this intrinsic coupling can lead to encoded universality, we 
analyze its action on three qubits, assuming for simplicity that the 
there is no dependence of the amplitude $J$ on the qubit indices, {i.e.}, 
$J_{ij} =J$. 
 
The Lie algebra ${\cal L}_E$ generated by ${\bf E}_{ij}$ on three qubits allows 
us to implement the Hamiltonians in the set $\{ {\bf E}_{12}, {\bf E}_{23}, 
{\bf E}_{13}, {\bf T} \equiv i( [{\bf E}_{12}, {\bf E}_{23}] \} $.  A better basis for the Lie algebra is given by the set of 
operators ${\bf H}_0={\bf E}_{12} +{\bf E}_{23} + {\bf E}_{13}$, ${\bf H}_1={1 
\over 4 \sqrt{3}} \left( {\bf E}_{13}- {\bf E}_{23} \right)$, ${\bf H}_3={1 
\over 12} \left( -2{\bf E}_{12} + {\bf E}_{23}+{\bf E}_{13} \right)$, ${\bf 
H}_2=i[{\bf H}_3, {\bf H}_1]$.  We then find that 
\begin{equation} 
[{\bf H}_0,{\bf H}_{\alpha}]=0 
\end{equation} 
for all $\alpha$ and 
\begin{equation} 
[{\bf H}_\alpha,{\bf H}_\beta]=i \epsilon_{\alpha \beta \gamma} {\bf H}_\gamma 
\label{eq:H_su2}
\end{equation} 
with $\alpha,\beta,\gamma \in \{1,2,3 \}$.  ${\bf H}_0$ is an abelian 
invariant   subalgebra of this Lie algebra and thus factors out as a 
global phase.  The set $\{ {\bf H}_1,{\bf H}_2,{\bf H}_3 \}$, on the 
other hand, act  as the generators of $su(2)$.  Thus the exchange 
interaction between three qubits can be used to implement a single 
encoded-qubit  $su(2)$. More precisely, we find that the ${\bf 
H}_\alpha$ for $\alpha \in  \{1,2,3 \}$ generates the algebra 
\begin{equation} 
{\cal L}_E^{(3)} =  \left({\cal L}_1 \otimes I_4 \right) \oplus \left( 
 {\cal L}_2 \otimes I_2 
\right) 
\end{equation} 
where ${\cal L}_d$ is the $d$-dimensional irrep of $su(2)$.  Note the 
degeneracy of the corresponding  irreps.  Corresponding to this 
decomposition the ${\bf H}_\alpha$ act as 
\begin{equation} 
{\bf H}_\alpha = {\bf 0}_4 \oplus \left(  {1 \over 2} { 
\sigma}_\alpha  \otimes {\bf I}_2\right) 
\end{equation} 
for $\alpha \in \{1,2,3\}$ where ${\bf 0}_4$ is 4-dimensional zero 
operator (the $1$ dimensional irreps all act as $0$) and ${\bf I}_2$ 
is the two-dimensional identity operator.  The action of the exchange 
is thus identical to that of an $su(2)$ operator on a single qubit 
when working over the encoded space defined by the above 
decomposition.  If we encode our logical qubits as 
\begin{eqnarray} \label{trio}
|0_L\rangle &=& {1 \over \sqrt{2}} (|010\rangle -|100\rangle) \nonumber 
\\ 
|1_L\rangle&=& \sqrt{2 \over 3} |001 \rangle -\sqrt{1 \over 6} |010 \rangle 
 - \sqrt{1 \over 6} |100 \rangle 
\end{eqnarray} 
then we find that the action of ${\bf H}_\alpha$ is ${1 \over 2}{ 
\sigma}_\alpha$.  Due to the degeneracy of the irrep, other encodings 
are also possible. Note that these states are nothing but $J=1/2$, 
$J_z=+1/2$  
total angular momentum states of $3$ spin-$1/2$ particles with a given 
projection along a certain axis, and can thus be found using 
elementary addition of angular momentum \cite{Kempe:00a}. 
 
Having shown that the action of the exchange interaction on three qubits can 
produce the effect of a single $2$-dimensional representation of $su(2)$, 
we now conjoin the qubits, i.e. 
induce a tensor product structure between blocks of three qubits in 
order to simulate a quantum circuit.  We thus {\em choose} an encoding 
scheme in which a single qubit is identified with $3$ physical qubits. 
This is not the only choice of tensor  product structure.  In fact, 
any tensor product between sets of $k \geq 3$  qubits can be used to 
construct a universal gate set\cite{Kempe:00a}. Once one has 
introduced a tensor structure for the encoding, it is  necessary to 
show that the natural couplings of the system can produce a 
non-trivial action between the tensor components of this  encoded 
tensor product structure, i.e., show that we get universal gates on 
the encoded qubits. 
 
In the above case, where we have shown that we can obtain full control 
($su(2)$) over an encoded qubit, what we now need to show 
is that a non-trivial action between the encoded qubits can be 
enacted.  This is nothing more than a map from the encoded universality 
to the fully universal set of physical gates consisting of single qubit 
gates supplemented by a non-trivial coupling between the qubits (like the $CNOT$, for example).  This 
point of the universality proof is typically the most daunting, but 
really amounts to nothing more than understanding the Lie algebra 
generated by the interaction over two tensored encoded qubits. 
 
For the exchange interaction we have shown explicitly that nearest neighbor 
interactions can be used to produce such an action 
\cite{Kempe:00a}.  In particular, we find that the effect of the 
exchange interaction on $6$ qubits becomes 
\begin{equation} 
{\cal L}_E^{(6)}= \left( {\cal L}_5 \otimes I_1 \right) \oplus \left( {\cal L}_9 \otimes I_3 \right) 
\oplus \left(  {\cal L}_5 \otimes I_5 \right) \oplus \left({\cal L}_1 \otimes I_7 \right)
\label{eq:L6e}
\end{equation} 
When we conjoin two encoded systems, the Hilbert space of the full 
system contains a subspace spanned by the two encoded systems.  It is 
the action on this conjoined subspace that we must then show can act 
nontrivially to produce an encoded two-qubit gate.  
When two of the above three-qubit codes are conjoined, we find that the 
resulting tensor product subspace
lie entirely within the second and third irreps of the decomposition, Eq.~(\ref{eq:L6e}).
This in turn implies that non-trivial 
interactions between states in the encoded irreps are possible.  In fact the 
original encoded single-qubit operations $su(2)$ are also contained within 
this 
decomposition.  The most important  point, however, is that on our 
established tensor product between  encoded qubits we can indeed 
couple these encoded qubits in such a way  as to achieve a map to the 
unencoded, fully universal set of gates.  Thus we have  established 
that with the exchange interaction, this encoding can efficiently (to within a factor of three in  spatial resources) simulate the unencoded set of universal gates. 
 
While the above example provides a systematic procedure to arrive at 
encoded universality, it is not unique.  Indeed, there is a great deal 
of arbitrariness in how to produce a universal  quantum computer using 
encoding.  Our scheme of taking small encodings and then coupling 
them by conjoining is by no means the only method to  obtain a universal quantum 
computer.  An interesting alternative approach that exploits 
topological aspects is also under active consideration, see e.g., 
\cite{Kitaev:97a,Preskill:97a}.  However, in dealing with real physical 
implementations, we expect that the use of small encodings will be very 
convenient.  The principle of conjoining small encoded qubits therefore 
provides a very useful and practical route. 

\subsection{Efficiency of Encoded Gate Implementation}
\label{sec:efficiency}
We now address the question of how efficiently these encoded gates can be
implemented with discrete operations.  Consider the encoded $su(2)$ defined
by the algebra of Eq.~(\ref{eq:H_su2}).  The Solovay-Kitaev theorem tells us 
that any encoded single-qubit operation can be implemented efficiently with
some sequence of discrete gates $e^{i{\bf H}t_n}$, where ${t_n}$ is the
set of discrete times over which the exchange operations are switched on.
However, it does not provide explicit values for ${t_n}$.  There does 
exist a brute force route to such sequences of discrete gates, namely via
truncation of the Trotter expansion, Eq.~(\ref{Trotter1}).  However there
is absolutely no guarantee that this expansion will be efficient.  

In recent
work\cite{DiVincenzo:00a} we have shown that while there does not yet appear to be an analytic
route to determination of optimally efficient discrete gate sequences,
it is nevertheless possible to use numerical methods to search for
efficient sequences.  Whether the resulting solutions are optimal or not
is not known, but in the example of isotropic exchange the numerical results
are striking.  In particular, the numerical search in Ref.~\cite{DiVincenzo:00a} showed that 
there exists a sequence of 19 discrete exchange gates that can achieve a nontrivial 
two-qubit gate between encoded blocks (the encoded CNOT), while
encoded one-qubit gates require no more than 4 elementary exchange interactions. 
Several numerical approaches to search for efficient
discrete gate sequences have been 
suggested\cite{Rains:97,Grassl:98,Burkard:99,Makhlin:00,Sanders:99}.  
Methods relying on local invariants of quantum states allow
determination of discrete gate sequences for two-qubit encoded gates
up to a one-qubit rotation\cite{Makhlin:00}.  This was the procedure 
employed in Ref.~\cite{DiVincenzo:00a} where two such invariants were
combined to yield a function whose minimal solution corresponds to a 
discrete gate sequence.  Clearly there are many other approaches to the
determination of optimal discrete gate sequences that could be explored,
including factors such as pulse shaping and genetic algorithms.

\subsection{A Useful Criterion for Demonstrating Non-Universality} \label{sec:criterion}
 
The above discussion has focused on how one can find universal encodings 
deriving from a given physical interaction. While we cannot always
guarantee that this is possible and in general will need to proceed on a 
case by case basis to find such encodings, formal conditions for the 
``inverse'' problem can nevertheless be made.   
Thus, in certain cases we can indeed answer the question of whether a 
set of interactions {\em cannot} be used as a universal quantum computer, 
even in the presence of arbitrary encoding.  We present here a useful 
criterion for testing this, allowing detection of Lie algebras which are 
not universal. 
 
Suppose one is given a set of Hamiltonians ${\tt H}_n$ which can be implemented 
in a Hamiltonian control sequence on $n$ subsystems.  Let ${\cal L}_n$ 
denote the Lie algebra which can be generated by ${\tt H}_n$, and let $g(n)$ 
denote the number of linearly independent operators in ${\cal L}_n$. We 
call $g(n)$ the subsystems growth function. 
\begin{theo} \label{th:growth} 
A growth function $g(n)$ which is polynomial in $n$ is not universal on a 
quantum circuit model. 
\end{theo} 
Sketch of the proof: The basic idea behind this theorem is the realization
that a quantum circuit 
model on $n$ subsystems has a state space which grows exponentially in $n$. 
Therefore performing unitary operators on this space is equivalent to 
generating elements of an exponentially growing Lie algebra. 
 
We will illustrate the above by giving a set of Hamiltonians which is not universal for {\em any} encoding of the information. 
Consider the set of gates generated by Hamiltonians in the set 
\begin{equation} 
{\cal O}^\prime= \{ {\sigma}_z^{(i)} , {\sigma}_x^{(i)} {\sigma}_x^{(i+1)} 
\} 
\end{equation} 
where $\sigma_\alpha^{(i)}$ is the $\alpha$th Pauli matrix operating on the 
$i$th qubit tensored with identity on all other qubits. We will now show that, 
even with the help of encoding, this set of Hamiltonians is not universal. 
 
Define the operation ${\bf M}_{jk,\alpha,\beta}= {\sigma}_\alpha^{(j)} 
\prod_{i=j+1}^{k-1} {\sigma}_z^{(i)} {\sigma}_\beta^{(k)}$ where $j<k$ and 
$\alpha,\beta \in \{x,y\}$.  We claim that the operators in the Lie algebra 
generated by ${\cal O}^\prime$ are all linear combinations of the form 
${\bf M}_{jk,\alpha,\beta}$ plus the single qubit ${\sigma}_z^{(i)}$. Notice 
that this is true for $n=2$.  We will prove the result by induction. First we 
note that because our generators are made up of Pauli operators, we need not 
worry about linear combinations of operators, but only need to worry about the 
operators which can be generated by commutation. Let ${\cal L}^n$ denote 
the Lie algebra on $n$ qubits generated by taking commutators in ${\cal 
O}^\prime$.  For example ${\cal L}_2=\{{\bf M}_{12,x,x},{\bf 
M}_{12,x,y},{\bf M}_{12,y,x},{\bf M}_{12,y,y},{\sigma}_z^{(1)}, 
{\sigma}_z^{(2)} \}$ as claimed above.  Assume that ${\cal L}_n= \{ {\bf 
M}_{jk,\alpha,\beta}, 1\leq j<k\leq n,\alpha,\beta \in \{x,y\} \} \cup \{ 
{\sigma}_z^{(i)}, 1\leq i \leq n \}$.  First notice that taking commutators of 
elements of ${\cal L}_n$ and ${\sigma}_z^{(i)}$ only produces elements in 
${\cal L}_n$: the only elements which ${\sigma}_z^{(i)}$ do not commute 
with are ${\bf M}_{ik,\alpha,\beta}$ and ${\bf M}_{ki,\alpha,\beta}$ and this 
commutation only serves to flip the value of $\alpha$ or $\beta$. Finally, note 
that taking the commutator between elements of ${\cal L}_n$ and 
${\sigma}_x^{(i)} {\sigma}_x^{(i+1)}$ can only generate elements that are linear combinations of the ${\bf M}_{jk,\alpha,\beta}$ and ${\sigma}_z^{(i)}$. 
To see this, first note that the only nontrivial commutators are 
those which occur with the ${\sigma}_z^{(i)}$ operators, which just produce 
elements in ${\cal L}_2$.  Further commutators between ${\sigma}_x^{(i)} 
{\sigma}_x^{(i+1)}$ and ${\bf M}_{jk,\alpha,\beta}$ only create ${\bf 
M}_{j^\prime k^\prime,\alpha^\prime,beta^\prime}$ which are one qubit larger or 
smaller. Thus we have proved that the Lie algebra generated by elements of 
${\cal O}^\prime$ is spanned by the set of linearly independent operators 
in ${\cal L}_n= \{ {\bf M}_{jk,\alpha,\beta}, 1\leq j<k\leq n,\alpha,\beta 
\in \{x,y\} \} \cup \{ {\sigma}_z^{(i)}, 1\leq i \leq n \}$. 
 
Let us count the operators in ${\cal L}_n$.  There are $n$ $\sigma_z^{(i)}$ 
operators and $4 {n \choose 2}$ ${\bf M}_{jk,\alpha,\beta}$ operators.  Thus 
the growth function for this Lie algebra is $g(n)=n+4 \left( {n(n-1) \over 2} 
\right) = 2n^2-n$.  This growth function is polynomial in $n$ and thus via 
Theorem \ref{th:growth} this set  of operators is not universal.

\section{Anisotropic Exchange} 
\label{anisotropic} 
 
Anisotropic Heisenberg spin couplings arise whenever there is some 
preferred direction in space along which the coupling is stronger or 
weaker. This could be due to, e.g., asymmetries induced by donor atoms 
in solid-state arrays of atoms coupled via their nuclear spins 
\cite{Kane:98,Kane:00}.  The XY-interaction arises when there is no 
coupling in the $z$-direction of the spins while the coupling in $x$- 
and $y$-direction is equally strong: 
\begin{eqnarray} 
({\bf H}_{XY})_{ij}  = \frac{J_{ij}}{2} ({  \sigma}_x^i {  \sigma}_x^j+{ 
\sigma}_y^i {  \sigma}_y^j) \equiv J_{ij} A_{ij}. 
\end{eqnarray} 
This situation is relevant to the proposal of solid-state quantum computation 
using quantum dot spins and cavity QED \cite{Imamoglu:99}. 
 
We will now analyze the power of the XY-interaction as a universal gate 
via encoding.  Several recent works have addressed ways in which the 
XY-interaction can be made universal under the addition of either
single-qubit operations\cite{Imamoglu:99,Wu:01a}, or single-qubit 
interactions\cite{Lidar:01a}.  In contrast to these proposals, our focus 
here is on deriving universality without any additional single-qubit 
interactions or operations.  
Our presentation here is designed not only to illustrate the 
general formalism and 
methodology of proof, but also to make the approach to find encodings
that provide universality with a {\em single} physical interaction,
transparent and applicable 
to other types of interactions (see also \cite{Kempe:ph}). Technical 
details of the proofs are relegated to the appendix \ref{App:A}. 
 
\subsection{Lie Algebra for Anisotropic Exchange} 
 
The set $\sf H=\{A_{ij}:\quad 1 \leq i < j \leq n\}$ generates the Lie-algebra 
$\cal L$. To help us find the splitting into irreps Eq. 
(\ref{eq:Liesplit2}) let us identify (by inspection) some elements in its 
commutant ${\cal L}^\prime$: \be \label{good} S_z  = \sum_{i=1}^n 
\sigma_z^i \quad \quad \quad {\bf X}  = \Pi_{i=1}^n \sigma_x^i \ee We claim 
that the associative algebra $\cal M$ generated by these two elements via 
linear combination and multiplication is identical to the commutant ${\cal 
L}^\prime$ of ${\cal L}$. Note that the commutant is always closed 
under linear combination and multiplication and hence is an associative 
algebra. Together with the identity operator, $\cal M$ obviously 
constitutes a $\dagger$ closed algebra 
and hence satisfies the condition of Theorem \ref{th:daggertheorem}. It splits 
into irreps according to Eq.~(\ref{eq:algebrasplit}). Assume for now that 
${\cal L}^\prime={\cal M}$. Then the splitting of ${\cal L}$ 
and the splitting of ${\cal M}$ (as well as that of its commutant ${\cal M}^\prime$) 
will be dual to each other: \be \label{eq:three} {\cal L} \cong \bigoplus_{J 
\in {\cal J}} {\cal{L}}_{J}(n_J) \otimes I_{d_J} \quad  {\cal M} 
\cong \bigoplus_{J \in {\cal J}}   I_{n_J} \otimes M(\CC^{d_J}) \quad 
{\cal M}^\prime  \cong \bigoplus_{J \in {\cal J}}   M(\CC^{n_J}) 
\otimes I_{d_J} \ee in the {\em same basis}. In particular the splitting of 
${\cal M}$ would give us the {\em encoding} into subspaces over which $\sf 
H$ is possibly a universal set. Note that to show that $\sf H$ is indeed universal over these subspaces we still have to prove $su(n_J) 
\subseteq {\cal{L}}_{J}(n_J)$.
 
Now consider that our claim of equality between ${\cal L}^\prime$ and
${\cal M}$ was wrong, and that actually ${\cal L}^\prime \supset {\cal 
M}$ holds. This implies that ${\cal L} \subset {\cal M}^\prime$. This 
would mean that the splitting of ${\cal M}^\prime$ is  {\em coarser} than 
the splitting of $\cal L$, since adding elements to $\cal L$ just 
``joins'' several previously separate factors $\bigoplus \CC^{n_J} \otimes 
\CC^{d_J}$ into one bigger block $\CC^{n_Jd_J} \otimes I$. This in turn would 
mean that $\cal L$ in this basis would have a sub-block structure. In 
particular, $\cal L$ could then not generate a full $su(n_J)$ in this 
block because it 
doesn't mix certain parts of the bigger block. 
 
Our strategy is then the following. We will first identify the splitting of $\cal 
M$ and its associated subspaces. Then, using an inductive proof, we shall 
show that over 
these subspaces $\cal L$ contains $su(n_J)$, thus allowing for universal 
computation. As a by product this will also prove our claim above that ${\cal 
L}^\prime={\cal M}$ does indeed hold. 
 
\subsection{The Algebra ${\cal M}$}

Let us study the algebra generated by the two elements $S_z$ and ${\bf X}$ in 
Eq. (\ref{good}).  They do not commute so we expect some irreducible 
representations of higher dimension. The element $S_z$ alone is already 
diagonal in the standard basis, the eigenvalue of a bit-string with $i$ ones 
and $n-i$ zeros is $n-2i$. The element ${\bf X}$ turns every zero into a one 
and vice versa. Let us rearrange all bit-strings into the following order 
\begin{equation} 
{\cal B}=\{000 \ldots 0, 111 \ldots 1, 100 \ldots 0, 011 \ldots 1, 010 
\ldots 0, 101 \ldots 1, \ldots \} 
\end{equation} 
i.e. listing elements with increasing number of ones and each followed by 
their transform under ${\bf X}$. In this basis $S_z$ and ${\bf X}$ are block-diagonal with 2-by-2 
blocks of the form $(n-2i) \sigma_z$ (for $S_z$) and $\sigma_x$ (for 
${\bf X}$), where $i=0 \ldots \lfloor n/2 \rfloor$ is the number of ones in the corresponding first 
basis bit-string of the particular block. Obviously all blocks with the same $i$ are the same, and there  are ${n \choose i}$ of these. By linear combination and multiplication we get the whole matrix algebra $M(\CC^2)$ on each block\footnote{There is one exception in the case of {\em even} $n$. Over all the ${n \choose n/2}$ states with equal number of zeros and ones $S_z$ acts as $0$. This means that over this subspace the algebra $\cal M$ is generated by ${\bf X}$ only, is thus {\em Abelian} and splits into one-dimensional irreps in the basis where ${\bf X}$ is diagonal. It is easy to see that this basis is formed by states $|s\ra + {\bf X}|s\ra$ (where ${\bf X}$ acts as $1$) and $|s\ra - {\bf X}|s\ra$ (where ${\bf X}$ acts as $-1$), where $|s\ra$ is a bit-string.  Over this subspace $\cal M$ splits into two different $1$-dimensional irreps with degeneracy ${n \choose n/2}/2$ as ${\cal M} \cong I_{{n \choose n/2}/2} \otimes M(\CC^1) \oplus I_{{n \choose n/2}/2} \otimes M(\CC^1)$.}. In the representation of $\cal M$ (\ref{eq:three}) we have ${\cal J}=\{0,1,2, \ldots \lfloor n/2 \rfloor\}$, $d_J=2$ and $n_J={n \choose J}$\footnote{and $n_{n/2}={n \choose n/2}/2$ for even $n$}. 
 
The largest encoding - assuming we can prove universality of $\cal L$ on these spaces - will be of dimension $n_J={n \choose (n-1)/2}$ for odd 
$n$ and ${n \choose {n/2-1}}$ for even $n$.\footnote{note: ${n \choose 
    {n/2}}/2 < {n \choose {(n/2-1)}}$ for $n>1$} 
The dimensions of the first few encodings are given as follows: 
\begin{center} 
\begin{tabular}{|c|c|c|c|c|c|c|c|c|} \hline 
n & 1 & 2 & 3 & 4 & 5 & 6 & 7 & 8 \\ \hline  max $n_J$ & 1 & 1 & 3 & 4 & 10 & 15 
& 35 & 56 \\ \hline 
\end{tabular} 
\end{center} 
Thus, for $n=3$ we can encode one qutrit, for $n=4$ we can encode two qubits, 
for $n=5$ one can either encode a single ten-level system, or two 
five-level
systems, or one can use only eight of the ten available states to
encode just three qubits, etc. 
As $n$ gets large, the rate of encoding approaches one (i.e. $\log_2 \max n_J \rightarrow n$), i.e. the encoding approaches unit efficiency. 
 
\subsection{Example - One Qutrit Encoded into Three Qubits}\label{sec:qutrit} 
 
The smallest space into which we can encode more than one state is the space of 
$n=3$ qubits. This encodes a logical {\em qutrit}. The algebra $\cal M$ 
and its commutant ${\cal L} \subset {\cal M}^\prime$ split as \be 
{\cal M} \cong I_3 \otimes M(\CC^2) \oplus I_1 \otimes M(\CC^2) \quad \quad 
{\cal M}^\prime \cong M(\CC^3) \otimes I_2 \oplus M(\CC^1) \otimes I_2. \ee 
We will show ``by hand'' that $\cal L$ generates all of $su(3)$ over the 
second product. According to the previous section we can chose the basis for 
our encoded qutrit as 
\begin{equation} 
\label{eq:qutrit} |0_L\rangle = |100\rangle \quad |1_L\rangle = |010\rangle \quad 
 |2_L\rangle = |001\rangle. 
\end{equation} 
The explicit action of $\sf H$ on this encoded space is 
\begin{eqnarray} 
A_{12}^L&=& 
\left( 
\begin{array}{ccc} 
  0 & 1 & 0 \\ 
  1 & 0 & 0 \\ 
  0 & 0 & 0 
\end{array} \right), \quad 
A_{23}^L= 
\left( \begin{array}{ccc} 
 0 & 0 & 0  \\  0 & 0 & 1 \\  0 & 1 & 0 
\end{array} \right), \quad 
A_{13}^L = 
\left( \begin{array}{ccc} 
 0 & 0 & 1 \\ 0 & 0 & 0 \\ 1 & 0 & 0 
\end{array} \right) , 
\end{eqnarray} 
where $A_{ij}^L$ is an operator on the encoded space.
Taking commutators of two of these encoded operators gives us $i \sigma_y$ on each two-dimensional subspace. Further commutators can easily be seen to
generate all of $su(3)$. Hence $\sf H$ is a universal set on the encoded qutrit. 
 
We can now introduce conjoining of the encoded qutrits (see Sec. \ref{tensor})
in order to make use of them as basic units in an encoded circuit. As 
outlined above, for this to be useful, we have to show that we can implement 
some non-trivial coupling between the qutrits such that 
 the full $su(3 \times 3)=su(9)$ can be generated between two conjoined blocks.  For example, it can easily be 
calculated that for the $6$ qubits made by conjoining the two $3$-state codes, the operator $[[A_{16},A_{15}],A_{12}]$ produces a coupling which preserves the conjoined coding {\it and} non-trivially couples the two encoded qutrits. By putting together encoded qutrits (conjoining) in this manner, the XY model can 
therefore serve as a universal quantum computer. 
 
Note that in general gates between encoded blocks do not constitute a 
problem once we have proved universality of $\sf H$ for higher dimensions. 
The $9$ basis states obtained by conjoining the $2$ encoded qutrits give states with exactly $2$ ones (and $7$ zeros).  
In  the $6$-qubit space they are all in the same subspace, namely the space 
generated by states with exactly $2$ ones. They span a $9$ dimensional 
subspace of this $15$ dimensional space. The general proof (see below and 
Appendix) will show that it is possible to implement $su(15)$ on this space, 
and in particular the sub-algebra $su(9) \oplus su(6) \subset su(15)$, 
where $su(9)$ acts on the $9$ states obtained by conjoining two encoded qutrits. 
Hence we can obtain any encoded gate between them. 
 
\subsection{General Proof} 
 
We will now proceed to prove that for $n$ qubits, where ${\cal M}^\prime$ splits as\footnote{the last component in the even case is $M(\CC^{{n \choose n/2}/2}) \otimes I_1 \oplus M(\CC^{{n \choose n/2}/2}) \otimes I_1$} 
\be 
{\cal M}^\prime \cong \bigoplus_{J \in \{0,1,\ldots \lfloor n/2 \rfloor\}} M(\CC^{n \choose J}) \otimes I_2 
\ee 
the Lie-algebra $\cal L$ contains $su(n_J)=su({n \choose J})$ on each 
component. Let us call each of the associated spaces $S_n(J)$. For instance 
$S_3(1)$ is spanned by the three states in Eq. (\ref{eq:qutrit}). In general 
we will make the canonical choice of taking our basis states to be those 
bit-strings which have $\#0 \geq \#1$\footnote{In the even $n$ case we will 
have two spaces $S^+_n(n/2)$ and  $S^-_n(n/2)$ of dimension 
${n \choose n/2}/2$, the first spanned by states of the form 
$|s\ra + {\bf X}|s\ra$ and the second by states of the form 
$|s\ra - {\bf X}|s\ra$.}. Then for the anisotropic exchange, the label $J$ 
actually counts the number of ones in the basis-bit-strings. 
 
We will proceed by induction, as in the corresponding proof for the exchange 
interaction \cite{Kempe:00a}, to show that on each space $\sf H$ generates 
$su(n_J)$ {\em independently} on $S_n(n_J)$, i.e. acting trivially on the rest of the space. We have seen in the previous section that in the case of $n=3$ $\sf H$ generates an {\em independent} $su(3)$ on $S_3(1)$ (since all three generators annihilate $S_3(0)=|000\ra$). Let us now understand how we build bigger spaces from smaller ones when we add a new qubit. To the basis states in $S_{n-1}(J)$ we can either add $0$ or $1$, which gives basis states in $S_n(J)$ and $S_n(J+1)$ respectively. Turning this around this means that states in $S_n(J)$ come either from $S_{n-1}(J)$ - if a $0$ was added - or from $S_{n-1}(J-1)$ - if a $1$ was added\footnote{The only exception are the spaces $S^\pm_{n}(n/2)$ for even $n$. They can only be built by {\em adding}  a $1$ as the $n$th qubit to states in $S_{n-1}(n/2-1)$. This gives us a set of ${n \choose n/2}/2$ strings $|s\ra$. We then form the linear combinations $|s\ra \pm {\bf X} |s\ra$ to obtain $S^{\pm}_n(n/2)$. We call this phenomenon {\em ``doubling''}. Similarly, to build basis states in $S_{n+1}(n/2)$ from $S^{\pm}_n(n/2)$ we have to take all the ${n \choose n/2}$ bit-strings with $n/2$ ones and add a $0$. We call this procedure {\em ``rearranging''}. {\em ``Doubling''} and {\em ``rearranging''} will require a more careful separate analysis in the inductive proof. }. 
 
For the inductive step let us assume that we can generate $su(n_J)$ {\em independently} on each subspace $S_{n-1}(J)$ of $(n-1)$ qubits. Within each space $S_n(J')$ of $n$ qubits, we shall call ${\bf 0}$ those states that come from 
$S_{n-1}(J')$ by adding a zero, and ${\bf 1}$ states that come from 
$S_{n-1}(J'-1)$ by adding a one. We call the states derived from going back 
two such steps ${\bf 00}$, ${\bf 01}$, ${\bf 10}$ and ${\bf 11}$ states, by 
analogy to the $TT$, $TB$, $BT$ and $BB$ states used in the proof for isotropic exchange \cite{Kempe:00a}. 
 
Now let us analyze how a $su$ on $S_{n-1}(J-1)$ affects the spaces of $n$ qubits\footnote{By this we mean a particular operation in $su(n_J)$ performed independently on this space.}. It propagates to a $su$ on the ${\bf 0}$-states of $S_n(J-1)$ and the same $su$ on the ${\bf 1}$-states of $S_n(J)$. In the first step we are going to show how to eliminate the action of this $su$ on {\em one} of these two spaces, keeping just the other one ({\em independence step}). This will give us an independent $su$ on the ${\bf 0}$ states of each $S_n$ and another independent $su$ on the ${\bf 1}$ states. In a second step ({\em mixing step}) we will show how to ``mix'' these two $su$ to obtain a full independent $su(n_J)$ on $S_n(J)$. These two steps can be found in the appendix \ref{App:A} which terminates the proof of universality of the XY-interaction on encoded spaces. 
 
\subsection{Conjoining} 
\label{conjoining}
 As we have seen on the example of the encoded {\em qutrit} in Sec. \ref{sec:qutrit} it is possible to obtain all encoded operations {\em between} the blocks, which we have to introduce to obtain a tensor-structure. Assume we cut our blocks of size $n$ qubits to encode a qu{\em dit}. The XY-interaction generates any operation within a block. If we join two blocks, its basis states will give a subset of the basis-bit-strings in one of the encoded spaces over $2n$ qubits. (In fact, if we used the space $S_n(J)$ over the blocks then the bigger block will be a subspace of $S_{2n}(2J)$.) We can universally compute over this conjoined block and in particular obtain any gate in the tensor space of our encoded qudits.\footnote{If we used even $n$ and one of the spaces 
$S^\pm(n/2)$ for our encoded qudit, we have to apply a little more care with 
the argument. However the conjoined space will be formed by states in either of $S_{2n}(n)$ and hence there is no problem.} 
Hence we can in principle choose any cut-off we wish and increase our coding efficiency as much as we like. 
 
We now make some remarks on efficiency for the encoded XY.  It is not clear 
how much overhead in time is required for actual implementation of
this encoded universality with the XY-interaction. Numerical
optimization studies similar to those we have made for the isotropic
exchange interaction in \cite{DiVincenzo:00a} and summarized in 
Section~\ref{sec:efficiency} have to be 
performed in order to find the minimal lengths of physical exchange gate 
sequences that can realize
specific gates between encoded blocks.
We note that several authors have recently shown that 
in a linear array, the XY-interaction 
is
not universal when restricted to act between nearest 
neighbors\cite{Valiant:01,Terhal:01a,Knill:01,Wu:01b} only. In fact this 
result can easily be derived using the criterion for non-universality of Sec. \ref{sec:criterion} by explicitly computing all possible commutators of nearest neighbor XY-exchanges. This set turns out to be polynomial in the number of qubits. This result implies that the {\em geometry} of allowed 
interactions for our encoded universality result with the XY-interaction will 
be important in construction of numerical gate sequences. 

\section{Error Correction and Leakage} \label{leakage}

We give here a brief discussion of the issues involved in making
encoded universal computation fault tolerant.  We do not pretend to
give a complete treatment of this subject, but we provide some
observations here that indicate that there is no fundamental obstacle
to making our coded computation fully fault tolerant.

Once we have a physical embodiment of a logical qubit (e.g., the sets
of three spins with the states given in Eq. (\ref{trio})), and a
procedure for implementing CNOT and one-qubit gates on these
qubits\cite{DiVincenzo:00a}, then all the fault tolerant procedures
developed recently for the standard gate model for quantum computation
can be applied\cite{Preskill:97a}.  The DFS coding would be concatenated
within a stabilizer code, the appropriate transversal implementation
of coded gates would be used, and appropriate error detection and correction
would be implemented that limits uncontrolled error propagation.  All
of this carries over unchanged in the DFS setting.

However, there is one aspect of error correction for which the DFS
setting is quite different, and that is in the issue of leakage.
Leakage is possible because the Hilbert spaces employed in quantum
computation are larger than is assumed in the ideal treatment.
Ideally, the state of a qubit lives in a two-dimensional Hilbert space
spanned by some states $|0_L\rangle$ and $|1_L\rangle$, and the
complete Hilbert space is just a tensor product of such Hilbert
spaces.  But for many physical embodiments of a qubit, the actual
dimension of its Hilbert space is larger, even though we may have
effective ways of restricting it approximately to just two of these
dimensions.  For example, an atom may have two low-lying energy levels
that we identify as the $0_L$ and $1_L$ states; but of course the atom
will have other excited states that we might label $|2_L\rangle$,
$|3_L\rangle$, etc.  Unintended interactions and imperfections will
surely, with some probability, bring the state of the qubit into these
extra states, although we hope that the probability of this occurring
is small.  This straying into extra dimensions is what is referred to
as leakage.

Leakage is actually a fairly likely error in the DFS schemes.  For
example, for the three-spin encoding of Eq. (\ref{trio}), The three
spins naturally have an 8-dimensional Hilbert space, of which we are
singling out two dimensions with quantum numbers $J=1/2$, $J_z=+1/2$.
But very simple kinds of unintended interactions, such as a magnetic
field affecting just one of the three spins, will cause the qubit to
``leak'' into the other six dimensions (those with quantum numbers
$J=1/2$, $J_z=-1/2$, or with $J=3/2$).  So it is important to
prescribe a scheme for detecting and correcting leakage to have a
fully fault tolerant scheme.

\begin{figure}
\begin{center}
\centerline{\mbox{\epsfig{file=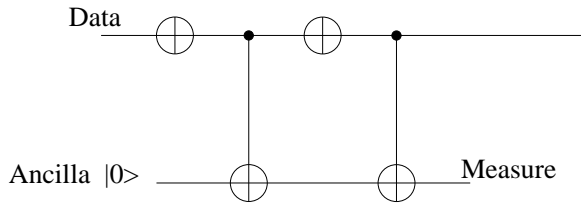,width=3in}}}
\vspace*{0.4cm}
\caption{Circuit suggested by Preskill for leakage detection.
\label{fig:leak1}}
\end{center}
\end{figure}

Preskill \cite{Preskill:97a} has indicated a simple procedure for
dealing with leakage.  He suggests that, from time to time, each qubit
be run through the simple circuit of Fig. \ref{fig:leak1}.  The truth
table of this circuit is to be
\begin{equation}
\begin{array}{ll}
&|0_L\rangle|0_L\rangle\rightarrow|0_L\rangle|1_L\rangle\\
&|1_L\rangle|0_L\rangle\rightarrow|0_L\rangle|1_L\rangle\\
&|2_L\rangle|0_L\rangle\rightarrow|2_L\rangle|0_L\rangle\\
&...\\
&|\mbox{leak}\rangle|0_L\rangle\rightarrow|\mbox{leak}\rangle|0_L\rangle.
\end{array}
\label{truth1}
\end{equation}
At the end of the circuit, the ancilla qubit is to be measured.  If it
is found to be a ``1'', the data qubit is in a valid state and it is
unchanged by the circuit.  But if it is found to be a ``0'', the data
qubit is known to be in a ``leaked'' state.  In this case, the data
qubit is replaced by a fresh $|0_L\rangle$ state.  This is usually
incorrect, but it is an ordinary bit error, and is dealt with by
the error-correction machinery at a higher level.

The general strategy introduced by Preskill can still be used for the
DFS-coded qubit, but the detailed procedure that he introduced will
not work.  From this point on we will specialize our remarks to the
case of coded computation using isotropic Heisenberg exchange.  The
problem is that the circuit given in Fig. \ref{fig:leak1} only has the
truth table given in Eq. (\ref{truth1}) given a certain assumption
about how the quantum gates in the circuit work.  In particular, it is
assumed, for both the NOT and the CNOT, that if one of the inputs is
leaky, the gate leaves the state unchanged:
\begin{equation}
\begin{array}{ll}
&\mbox{assumed CNOT action in Fig. \protect\ref{fig:leak1}:}\\
&|0_L\rangle|0_L\rangle\rightarrow|0_L\rangle|0_L\rangle\\
&|1_L\rangle|0_L\rangle\rightarrow|1_L\rangle|1_L\rangle\\
&|2_L\rangle|0_L\rangle\rightarrow|2_L\rangle|0_L\rangle\\
&|3_L\rangle|0_L\rangle\rightarrow|3_L\rangle|0_L\rangle\\
&... .
\end{array}
\label{truthCNOT}
\end{equation}
We find that it is not possible to implement a coded CNOT using the
Heisenberg exchange that has this extended truth table, as it would
require the gate to change the total spin quantum number of the state.
Therefore, we must search for a different approach to leakage
detection and correction.

\begin{figure}
\begin{center}
\centerline{\mbox{\epsfig{file=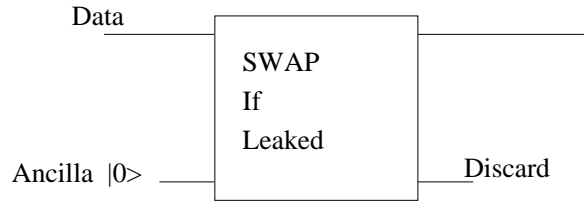,width=3in}}}
\vspace*{0.4cm}
\caption{Another leakage repair procedure better adapted to coded computation.}
\label{fig:leak2}
\end{center}
\end{figure}

Fortunately, we find that a modified leakage-correction circuit
specification is permitted by the constraints of the exchange
interaction; this circuit can then be used in the same way as the
Preskill leakage-correction circuit within a complete fault-tolerant
procedure.  In fact, the circuit action we achieve, indicated
symbolically in Fig. \ref{fig:leak2}, can be used in a simpler
way in fault tolerant quantum computation:
the leakage is corrected without the need for any
measurement.  What we mean by ``SWAP If Leaked'' ({\em SIL}) in
Fig. \ref{fig:leak2} is the following truth table:
\begin{equation}
\begin{array}{ll}
&\mbox{``SWAP If Leaked'', Fig. \protect\ref{fig:leak2}:}\\
&|0_L\rangle|0_L\rangle\rightarrow|0_L\rangle|0_L\rangle\\
&|1_L\rangle|0_L\rangle\rightarrow|1_L\rangle|0_L\rangle\\
&|2_L\rangle|0_L\rangle\rightarrow|0_L\rangle|2_L\rangle\\
&|3_L\rangle|0_L\rangle\rightarrow|0_L\rangle|3_L\rangle\\
&... .
\end{array}
\label{truthSIL}
\end{equation}
We note that if the data qubit is not leaked, it is not disturbed; if
it is leaked, it is always replaced by $|0_L\rangle$, permitting
subsequent correction.  (Of course, there is now no definite signal of
when correction must be done.  But if the rate of leakage is known,
then error correction can be done with the appropriate frequency.)

Actually, the function of ``SWAP If Leaked'' is achieved with a more
relaxed specification:
\begin{equation}
\begin{array}{ll}
&\mbox{modified ``swap if leaked'':}\\
&|0_L\rangle|0_L\rangle\rightarrow|0_L\rangle|0_L\rangle\\
&|1_L\rangle|0_L\rangle\rightarrow|1_L\rangle|0_L\rangle\\
&|\mbox{leak}\rangle|0_L\rangle\rightarrow|\mbox{anything in 
$0_L$-$1_L$ space}\rangle|X\rangle.
\end{array}
\label{truthSIL2}
\end{equation}
Here $|X\rangle$ is any state vector consistent with unitarity.
This generalization will be needed in our DFS analysis.

Now, we consider the implementation of {\em SIL} using the three-spin
DFS coding discussed earlier.  For this we adopt a specific notation
for these states that will make the discussion easier:
\begin{equation}
|N,n,J,J_z\rangle.
\end{equation}
Here $N$ is the number of spins in the DFS (which is not really a
quantum number, just a useful label), $J$ is the total angular
momentum, $J_z$ is its projection on the $z$ axis, and $n$ is a
``principal'' quantum number that labels the different states in the
DFS (e.g., $n$ runs from 0 to 8 for the nine triplet states with $N=6$,
$J=1$).

In this notation our two coded qubit states are
\begin{equation}
|0_L\rangle=|3,0,1/2,+1/2\rangle,\ \ |1_L\rangle=|3,1,1/2,+1/2\rangle
\end{equation}
We will assume that a three-spin ancilla is available that has been
freshly prepared in the coded 0 state.  Thus, we need to consider the
operation of the {\em SIL} circuit on the eight basis states (the
ancilla state is second):
\begin{equation}
\begin{array}{ll}
1.&|3,0,1/2,+1/2\rangle|3,0,1/2,+1/2\rangle\\
2.&|3,1,1/2,+1/2\rangle|3,0,1/2,+1/2\rangle\\
3.&|3,0,1/2,-1/2\rangle|3,0,1/2,+1/2\rangle\\
4.&|3,1,1/2,-1/2\rangle|3,0,1/2,+1/2\rangle\\
5.&|3,0,3/2,+3/2\rangle|3,0,1/2,+1/2\rangle\\
6.&|3,0,3/2,+1/2\rangle|3,0,1/2,+1/2\rangle\\
7.&|3,0,3/2,-1/2\rangle|3,0,1/2,+1/2\rangle\\
8.&|3,0,3/2,-3/2\rangle|3,0,1/2,+1/2\rangle.
\end{array}
\end{equation}
3. through 8. are the ``leaked'' cases.  To impose the
constraint that the {\em SIL} circuit be implemented only with isotropic
Heisenberg interactions, we must write these eight states in the basis
of total angular momentum states for the six-spin block:
\begin{equation}
\begin{array}{ll}
1.&|3,0,1/2,+1/2\rangle|3,0,1/2,+1/2\rangle=|6,0,1,1\rangle\\
2.&|3,1,1/2,+1/2\rangle|3,0,1/2,+1/2\rangle=|6,1,1,1\rangle\\
3.&|3,0,1/2,-1/2\rangle|3,0,1/2,+1/2\rangle=1/\sqrt{2}|6,0,0,0\rangle+1/\sqrt{2}|6,0,1,0\rangle\\
4.&|3,1,1/2,-1/2\rangle|3,0,1/2,+1/2\rangle=1/\sqrt{2}|6,1,0,0\rangle+1/\sqrt{2}|6,1,1,0\rangle\\
5.&|3,0,3/2,+3/2\rangle|3,0,1/2,+1/2\rangle=|6,0,2,2\rangle\\
6.&|3,0,3/2,+1/2\rangle|3,0,1/2,+1/2\rangle=a|6,4,1,1\rangle+b|6,0,2,1\rangle\\
7.&|3,0,3/2,-1/2\rangle|3,0,3/2,+1/2\rangle=c|6,4,1,0\rangle+d|6,0,2,0\rangle\\
8.&|3,0,3/2,-3/2\rangle|3,0,3/2,+1/2\rangle=e|6,4,1,-1\rangle+f|6,0,2,-1\rangle.
\end{array}
\end{equation}
{\em a-f} are Clebsch-Gordan coefficients whose values we will not need.
Recall that in this basis, the Heisenberg interactions cannot change
the $J$ quantum number, and the action of the circuit must be
independent of the $J_z$ quantum number.

The ``no-leak'' cases 1. and 2. require, from Eq. (\ref{truthSIL}):
\begin{equation}
\begin{array}{l}
SIL|6,0,1,1\rangle=|6,0,1,1\rangle\\
SIL|6,1,1,1\rangle=|6,1,1,1\rangle.
\end{array}
\end{equation}
But because of the $J_z$ invariance of any circuit {\em SIL} made up only 
of
exchanges, this implies
\begin{equation}
\begin{array}{l}
SIL|6,0,1,S_z\rangle=|6,0,1,S_z\rangle\\
SIL|6,1,1,S_z\rangle=|6,1,1,S_z\rangle.
\label{res1}
\end{array}
\end{equation}
This has implications for cases 3. and 4.  We can use the decompositions
\begin{equation}
\begin{array}{l}
|6,0,1,0\rangle=1/\sqrt{2}|3,0,1/2,-1/2\rangle|3,0,1/2,+1/2\rangle+
                1/\sqrt{2}|3,0,1/2,+1/2\rangle|3,0,1/2,-1/2\rangle\\
|6,1,1,0\rangle=1/\sqrt{2}|3,1,1/2,-1/2\rangle|3,0,1/2,+1/2\rangle+
                1/\sqrt{2}|3,1,1/2,+1/2\rangle|3,0,1/2,-1/2\rangle
\end{array}
\end{equation}
to write these out as
\begin{equation}
\begin{array}{l}
SIL|3,0,1/2,-1/2\rangle|3,0,1/2,+1/2\rangle=\\(1/\sqrt{2})SIL(|6,0,0,0\rangle)+
1/2|3,0,1/2,-1/2\rangle|3,0,1/2,+1/2\rangle+
1/2|3,0,1/2,+1/2\rangle|3,0,1/2,-1/2\rangle\\
SIL|3,1,1/2,-1/2\rangle|3,0,1/2,+1/2\rangle=\\(1/\sqrt{2})SIL(|6,1,0,0\rangle)+
1/2|3,1,1/2,-1/2\rangle|3,0,1/2,+1/2\rangle+
1/2|3,1,1/2,+1/2\rangle|3,0,1/2,-1/2\rangle.
\end{array}
\end{equation}
We see that if we make the choice
\begin{equation}
\begin{array}{l}
SIL|6,0,0,0\rangle=-|6,0,0,0\rangle\\
SIL|6,1,0,0\rangle=-|6,1,0,0\rangle
\label{res2}
\end{array}
\end{equation}
(note the sign change), then
using these and the relations
\begin{equation}
\begin{array}{l}
|6,0,0,0\rangle=1/\sqrt{2}|3,0,1/2,-1/2\rangle|3,0,1/2,+1/2\rangle-
                1/\sqrt{2}|3,0,1/2,+1/2\rangle|3,0,1/2,-1/2\rangle\\
|6,1,0,0\rangle=1/\sqrt{2}|3,1,1/2,-1/2\rangle|3,0,1/2,+1/2\rangle-
                1/\sqrt{2}|3,1,1/2,+1/2\rangle|3,0,1/2,-1/2\rangle,
\end{array}
\end{equation}
we can plug back in to get
\begin{equation}
\begin{array}{l}
SIL|3,0,1/2,-1/2\rangle|3,0,1/2,+1/2\rangle=
|3,0,1/2,+1/2\rangle|3,0,1/2,-1/2\rangle\\
SIL|3,1,1/2,-1/2\rangle|3,0,1/2,+1/2\rangle=
|3,1,1/2,+1/2\rangle|3,0,1/2,-1/2\rangle.
\end{array}
\end{equation}
Here we achieve the generalized ``SWAP If Leaked'' requirement of 
Eq. (\ref{truthSIL2}).

The last four cases are easier, because they are not constrained by
the earlier choices.  To deal with these, we introduce some new state
expansions:
\begin{equation}
\begin{array}{ll}
5.&|3,0,1/2,+1/2\rangle|3,0,3/2,+3/2\rangle=|6,2,2,2\rangle\\
6.&|3,0,1/2,+1/2\rangle|3,0,3/2,+1/2\rangle=a|6,7,1,1\rangle+b|6,2,2,1\rangle\\
7.&|3,0,1/2,+1/2\rangle|3,0,3/2,-1/2\rangle=c|6,7,1,0\rangle+d|6,2,2,0\rangle\\
8.&|3,0,1/2,+1/2\rangle|3,0,3/2,-3/2\rangle=e|6,7,1,-1\rangle+f|6,2,2,-1\rangle.
\end{array}
\end{equation}
Note that the Clebsch-Gordan coefficients {\em a-f} are the same as above.
Then, we can fix the action of {\em SIL} to be
\begin{equation}
\begin{array}{l}
SIL|6,0,2,J_z\rangle=|6,2,2,J_z\rangle\\
SIL|6,4,1,J_z\rangle=|6,7,1,J_z\rangle,
\label{res3}
\end{array}
\end{equation}
which will give us
\begin{equation}
\begin{array}{l}
SIL|3,0,3/2,+3/2\rangle|3,0,1/2,+1/2\rangle=|3,0,1/2,+1/2\rangle|3,0,3/2,+3/2\rangle\\
SIL|3,0,3/2,+1/2\rangle|3,0,1/2,+1/2\rangle=|3,0,1/2,+1/2\rangle|3,0,3/2,+1/2\rangle\\
SIL|3,0,3/2,-1/2\rangle|3,0,1/2,+1/2\rangle=|3,0,1/2,+1/2\rangle|3,0,3/2,-1/2\rangle\\
SIL|3,0,3/2,-3/2\rangle|3,0,1/2,+1/2\rangle=|3,0,1/2,+1/2\rangle|3,0,3/2,-3/2\rangle,
\end{array}
\end{equation}
as desired.  This completes the analysis.  We see that the
specifications of Eqs. (\ref{res1},\ref{res2},\ref{res3}) for the
desired circuit are for a unitary transformation that is obtainable by
only Heisenberg interactions, by the coded universality theorem of
\cite{Kempe:00a}.  We have not determined an explicit circuit for {\em
SIL}; it might be interesting to use the techniques of
\cite{DiVincenzo:00a} to find one.  Note that the specifications
Eqs. (\ref{res1},\ref{res2},\ref{res3}) for {\em SIL} only constrain
its action on 16 of the 64 dimensions of the Hilbert space; its action
on the remaining 48 dimensions is completely arbitrary, except for the
constraints of unitarity.

\section{Outlook and Open Questions}\label{outlook} 
 
It is unclear at this point which general kinds of physical interactions are amenable to ``encoded 
universality''. As we have seen, both the exchange and the XY-interaction give 
rise to a Lie-algebra that splits into degenerate irreducible representations. 
The degeneracy of this decomposition is the key to finding a suitable encoding. In 
physics, mathematical degeneracies are often linked to symmetries of a system, 
and indeed both the exchange and the XY-interaction possess a significant
amount of 
symmetry. (The isotropic exchange is invariant under permutation of the 
Pauli matrices, and 
the XY-interaction is invariant under permuting $\sigma_x$ and $\sigma_y$.) 
On 
the other hand, we know from previous work of Lloyd and others \cite{Lloyd:95a}
that each {\em generic} (and in particular 
non-symmetric) two-qubit interaction is universal {\em without} encoding if 
we allow 
bits to be flipped (i.e., the exchange gate). Note that both the exchange gate 
and the XY-interaction are invariant under the permutation of the specific
two qubits on which they act.
It would be interesting to know what the ``encoded power'' 
of other interactions not supplemented by an exchange interaction is, and also
how 
this power is related to their symmetry properties. 
 
Another interesting question is related to the topology and arrangement of the 
physical qubits of the system. In many cases only nearest-neighbor interactions can be enacted. 
In our formalism we have allowed for application of the gate between arbitrary 
pairs of qubits. Thus, it would be important to know whether a restricted set of 
gates, for instance, only between qubits $i$ and $i+1$, has the same 
universality power.
Clearly, in the case of the exchange interaction, we can 
limit ourselves to only nearest neighbor interactions, because we can
decompose all permutations into such pairs.  Thus exchange can be used to
flip qubits at our convenience.  However, as indicated above (Section~\ref{conjoining}), this is not true for the case of the XY-interaction. Whether for example certain two-dimensional layouts might be sufficient is still unclear. 
 
A third open question is concerned with the overhead in time, i.e., the number 
of gates involving physical interactions switched on for discrete time 
intervals that are needed to realize an ``encoded'' standard gate. 
As discussed in Section~\ref{sec:efficiency}, there is currently no analytic
route to finding the optimal gate sequences and we cannot even estimate
the length of this in the general case.  Establishing optimal discrete 
sequences is currently dependent on numerical searches in many dimensions.  
It would be highly
desirable to develop systematic bounds on the length of sequences, even if
an analytical route to optimization remains elusive.  
 
\vspace{0.5cm}
{\bf Acknowledgements:}
JK, DB, and KBW's effort is sponsored by the Defense Advanced  
Research Projects Agency (DARPA) and Air Force Laboratory, Air Force 
Materiel  
Command, USAF, under agreement number F30602-01-2-0524. The U.S. Government 
is  
authorized to reproduce and distribute reprints for Governmental purposes  
notwithstanding any copyright annotation thereon.  
The views and conclusions contained herein are those of the authors and
should not be interpreted as necessarily representing the official
policies or endorsements, either expressed or implied, of the Defense
Advanced Research Projects Agency (DARPA), the Air Force Laboratory, or
the US Government.
DPD's effort was supported by the National Security Agency (NSA) and the 
Advanced Research and Development Activity (ARDA) under Army Research Office (ARO) contract number DAAG55-98-C-0041.

\appendix 
\section{Details of the Universality Proof for the XY interaction}\label{App:A} 
 
\begin{quote} {\em Claim: Independence --} Using operations in the independent $su$ on $S_{n-1}(J-1)$ and $A_{n-1,n}$ we can generate an independent $su$ on the ${\bf 1}$-states of $S_n(J)$. 
\end{quote} 
{\em Proof:} Use two ${\bf 0}$ states in $S_{n-1}(J-1)$ and a $\sigma_z$ on them. This propagates to a $\sigma_z$ on the ${\bf 01}$-states of $S_n(J)$ and to a  $\sigma_z$ on the ${\bf 00}$-states of $S_n(J-1)$. Commute this $\sigma_z$ with $A_{n-1,n}$. Since $A_{n-1,n}$ annihilates ${\bf 00}$-states the net action on $S_n(J-1)$ will be zero, whereas the net action on $S_n(J)$ will be $i\sigma_y$ between one of the ${\bf 01}$-states and another ${\bf 10}$ state. Commute again with $A_{n-1,n}$ to obtain an {\em independent } $\sigma_z$ on the original two ${\bf 01}$-states in $S_n(J)$. This new $\sigma_z$, now entirely acting only on $S_n(J)$, can be used via commutation with operations in the $su$ to generate the now {\em independent} full $su$ on the ${\bf 1}$-states of $S_n(J)$ with the same arguments as in \cite{Kempe:00a} (using the Mixing Lemma of App. C there). 
 
When does this method not work? Only if there are less than $2$ ${\bf 
0}$-states in $S_{n-1}(J-1)$. If $n-1 \geq 3$ this is never the case except for 
$S_{n-1}(0)$, which just contains the $|00\ldots 0\ra$ state. In this case 
$S_n(1)$ just contains one ${\bf 1}$-state and $su(1)=0$ is trivially 
implemented on it\footnote{Let us treat the case of even $n$ here: {\em 
``Doubling'' --} If we can implement $su$ on $S_{n-1}(n/2-1)$ by construction 
this gives us coupled simultaneous $su$'s on the two spaces $S^\pm_n(n/2)$ 
together with the same $su$ on the ${\bf 0}$-states of $S_n(n/2-1)$. The action 
on these latter states can be eliminated as before and we are left with these 
two coupled full $su$'s on $S^\pm_n(n/2)$. We will not attempt to eliminate the 
action of one of them, for practical purposes, if we need to use these spaces, 
we will only be able to use one of them. {\em ``Rearranging'' --  } The 
problematic space is $S_{n+1}(n/2)$, more precisely its ${\bf 0}$-states. A 
priory we only obtain two coupled smaller $su$'s on this ${\bf 0}$-space, and 
not the full $su$. However since rearranging provides a change of basis we do 
not have a block-structure and it is easy to see that commuting with 
$A_{n-1,n}$ mixes the two $su$'s to give the full $su$ on the ${\bf 0}$-states 
of $S_{n+1}(n/2)$.}. 
 
Of course given independent $su$ on the ${\bf 1}$-states of $S_n$ we can 
immediately also get an independent $su$ on the ${\bf 0}$-states by subtracting 
the former from the coupled action of a $su$ on a $S_{n-1}$. 
 
\begin{quote}{\em Claim: Mixing --} Given an independent $su$ on each of the ${\bf 0}$-states resp. ${\bf 1}$-states of $S_n$ closing the Lie-algebra with $A_{n-1,n}$ gives the full $su$ on $S_n$ 
(independent). 
\end{quote} 
 
{\em Proof:} To get a $i \sigma_y$ between a ${\bf 0}$-state and a ${\bf 1}$ state take a $\sigma_z$ between two ${\bf 10}$ states (or between a ${\bf 00}$ and a ${\bf 10}$-state) and commute it with $A_{n-1,n}$. To get a $\sigma_z$ between these two states, commute with $A_{n-1,n}$ again. This gives a $su(2)$ connecting ${\bf 0}$- and ${\bf 1}$-states. The result follows from the Enlarging Lemma in \cite{Kempe:00a} App. C. This proof works for all sizes of $S_n$. 
\bibliographystyle{prsty} 

\end{document}